Nesting a Target Study within a Target Trial: A Framework for Evaluating Intervention Effects on Disparities

Xinyi Sun,[1] Theodore J. Iwashyna,[2,3] Emmanuel F. Drabo,[3] Deidra C. Crews,[2] Kadija Ferryman,[3,4] John W. Jackson[1,5,6,7,8]

1. Department of Epidemiology, Johns Hopkins Bloomberg School of Public Health, Baltimore, MD, United States
2. Department of Medicine, Johns Hopkins School of Medicine, Baltimore, MD, United States
3. Department of Health Policy and Management, Johns Hopkins Bloomberg School of Public Health, Baltimore, MD, United States
4. Berman Institute of Bioethics, Johns Hopkins University, Baltimore, MD, United States
5. Department of Biostatistics, Johns Hopkins Bloomberg School of Public Health, Baltimore, MD, United States
6. Department of Mental Health, Johns Hopkins Bloomberg School of Public Health, Baltimore, MD, United States
7. Johns Hopkins Center for Health Equity, Johns Hopkins University, Baltimore, MD, United States
8. Johns Hopkins Center for Health Disparities Solutions, Baltimore, MD, United States

**Author contributions**

JWJ conceived, organized, and supervised the work. XS contributed to the methodology development, developed the formal results, conducted the data application, and drafted the initial and revised manuscripts. TJI, EFD, DCC, and KF contributed to the methodology development, reviewed the data application, and offered expert guidance throughout the work. TJI, EFD, DCC, KF, and JWJ critically edited the initial and revised manuscripts for scientific content and clarity.

**Conflict of interest statement**

TJI reported receiving grants from the National Institutes of Health and serving as a consultant and potential expert witness for Roots Community Health on matters related to pulse oximetry, outside the submitted work. EFD reported receiving consultation honorarium from Pfizer outside the submitted work. DCC reported receiving grants from Somatus and Baxter outside the submitted work. KF reported receiving personal fees for serving on the institutional review board of the National Institutes of Health's All of Us Research Program and being a member of the Digital Ethics Advisory Panel for Merck KGaA. No other disclosures were reported.

**Funding**

This work was supported by a Nexus Research Award from Johns Hopkins University and by the Robert E. Meyerhoff Foundation.

**Acknowledgements**

We thank Trang Q. Nguyen for valuable input on missing data considerations and multiple imputation.

**Data availability statement**

The empirical data used in the study is not publicly shareable due to IRB restrictions. Sample code demonstrating implementation of the framework are available at: https://github.com/J-J-Lab/TSTT-sample-code.

**Abstract**

We present a novel framework (TS+TT) to nest a Target Study (TS) within a Target Trial (TT) for evaluating the effects of interventions on disparities. The TS component grounds the measurement of disparity in ethical assumptions, based on the concept of allowability, and anchors it to an explicit population within calendar time. It specifies an enrollment plan of stratified sampling of eligible persons to yield a sample where social groups are distributionally similar on covariates deemed allowable for measuring disparity. Within this enrolled sample, the TT component specifies randomization of intervention strategies within each social group. Because social groups are similarly situated on allowable covariates at baseline, and because assigned intervention arms are exchangeable within social groups, TS+TT reflects a meaningful causal estimand for evaluating how interventions impact disparity. We describe the framework's key components, its emulation, and demonstrate its application to evaluate how hypothetical interventions on pulse oximeter bias affect disparities in treatment receipt in clinical care. We also extend semiparametric G-computation to accommodate continuous stochastic interventions and estimate counterfactual disparities in time-to-event outcomes. The TS+TT framework offers a versatile and policy-relevant approach for generating ethically informed causal evidence to reduce disparities and avoid exacerbating disparities.

## 1. Introduction

Disparities in healthcare-related outcomes are defined by the Institute of Medicine (IOM) as "differences in the quality of care that are not due to access-related factors or clinical needs, preferences, and appropriateness of intervention",[1] a definition which has been influential.[2] While certain components of this definition (e.g., access, preferences) are debatable, it places ethics at the center of disparity measurement—highlighting the need to distinguish between fair and unfair sources of difference.[3] This principle, often referred to as *allowability*, builds normative and ethical assumptions into the disparity estimand. Allowable covariates that reflect fair sources of difference (if any) should be adjusted for when measuring a disparity, whereas non-allowable covariates that reflect social disadvantage and contribute to the disparities should not.[4] (See [3,5] for discussion on defining allowable and non-allowable covariates).

Despite the central importance of allowability, many analytic approaches used to evaluate the impact of interventions on disparities fail to incorporate this principle. Traditional regression estimators typically condition on all covariates required for confounding control or statistical efficiency, implicitly treating them as appropriate for defining disparity.[6] When these covariates include non-allowables (e.g., socioeconomic status (SES)), such adjustments can remove unfair components of the disparity that we aim to characterize and doing so can distort the magnitude and direction of the intervention effect on disparity.[6]

Meanwhile, the Target Trial (TT) framework provides a rigorous approach for estimating intervention effects on health outcomes using observational data.[7] By aligning study components to emulate a hypothetical randomized trial, TT clarifies the research question and estimand, mitigates common biases, and supports causal inference. However, the original TT framework was not designed to evaluate interventions in terms of their effects on outcome disparities. Although TT balances all potential confounders across intervention arms (to mimic randomization), it does not specify which covariates should be treated as allowable when defining disparities. If applied to the disparity setting, TT would implicitly treat all group differences as non-allowable, leaving social groups imbalanced on these factors. Such approaches are adequate when disparity can be conceptualized as inequality (see [8] for an example in epidemiology, [9] for an example in sociology, and [10] for further discussion of inequality vs. disparity). But they fall short in healthcare settings where differences in clinical need are often considered allowable and should be balanced across social groups, as outcome differences stemming from the differential distribution of allowables are often not considered morally concerning, and are thus excluded from measures of disparities in healthcare outcomes (e.g., receipt or timing of treatment).[4,11] Because the interpretation of disparity and intervention effect estimates depends on implicit normative choices, existing methods may have limited value for evaluations of fairness and for policy decision-making.

Recent work by Jackson and colleagues has begun to close this gap by reviewing allowability-based criteria[5] and developing analytic approaches for estimating intervention effects on disparities in both health outcomes and healthcare decision-making.[6] They also introduced the Target Study (TS) model to measure disparities descriptively while accounting for allowability, using a sampling-based design that can be emulated using observed data.[12]

In this paper, we build on these advances by proposing a unified framework that nests the Target Study within the Target Trial (henceforth referred to as TS+TT) to evaluate the effects of hypothetical interventions on disparity. Our TS+TT framework introduces two key innovations, extending the TT framework in two ways. First, we prespecify an enrollment window and a stratified sampling design that, together, reflect a meaningful disparity estimand that is grounded in ethical principles and anchored to a well-defined population in calendar time. Second, we conceptualize a stratified randomized trial in which randomization occurs within each social group, to ensure exchangeability across interventions and support causal inference. By embedding TS within TT, our framework enables researchers to estimate intervention effects on disparities in a principled and morally interpretable way. In our motivating example, we also adapt semiparametric G-computation to evaluate stochastic intervention effects on disparities in time-to-event outcomes. Together, this work offers a flexible, policy-relevant framework for both identifying and addressing disparities across outcome and intervention types.

## 2. Motivating example

Pulse oximeter (PO) estimates peripheral oxygen saturation ($SpO_2$), whereas true saturation is measured by arterial blood gas ($SaO_2$). Historically, $SpO_2$ overestimates $SaO_2$ in persons with dark skin, potentially leading to delayed treatment and worse health outcomes.[13] The FDA's 2025 draft guidance emphasizes the need to improve PO performance across diverse skin tones.[14] In healthcare data, race and ethnicity are often used as proxies for skin tone,[15] as direct measures are typically unavailable in electronic medical records (EMR). Let $G = 1$ denote the disadvantaged group (e.g., non-Hispanic Black [NHB]), and $G = 0$ the advantaged group (e.g., non-Hispanic White [NHW]). Suppose a healthcare regulator seeks to evaluate how differential PO accuracy by race impact disparities in clinical care, such as dexamethasone treatment in COVID-19 patients. These considerations are central to determine what levels of bias should be considered as "acceptable" for regulating device development, validation, and clinical use.

## 3. Specification of the TS+TT framework

### 3.1 Overview

Since the general structure of TT is well established,[7] and TS is formally introduced elsewhere[12] and demonstrated in a tutorial,[16] we focus on design features suited to evaluating interventions effects on disparities. **Table 1** gives a complete protocol for our motivating example.

We envision a multi-arm trial testing hypothetical PO devices, each designed with a prespecified degree of differential measurement error across racial groups. The population of interest is adults receiving inpatient care for COVID-19 at a large regional health system in the U.S. During each enrollment window indexed by calendar time $k$, we first sample eligible persons from each social group of interest (e.g., NHB and NHW) so that the distribution of allowable covariates $\boldsymbol{A_k}$ matches that of the within-study standard population $T = 1$ (e.g., NHB), achieving balance in the allowables across social groups. We then randomize persons within each group to intervention arms. Enrolled persons are followed for $J$ units of time for outcome $Y_{k+J}$.

The causal estimand of interest is the effect of an intervention strategy $\tilde{z}$ (e.g., a device with a certain degree of differential accuracy) on disparity at time $k$, compared to a reference strategy $\tilde{z}_0$ (e.g., a device with nondifferential accuracy).

For a given strategy $\tilde{z}$, the additive disparity at time $k$ is defined as the difference in average outcomes between the disadvantaged group ($G = 1$) and the advantaged group ($G = 0$):

$$\psi_k(\tilde{z}) = \mu_k^{\tilde{z}}(G = 1) - \mu_k^{\tilde{z}}(G = 0) \quad (1)$$

where $\mu_k^{\tilde{z}}(g) = \int E\left[Y_{k+J}^{\tilde{z}} \middle| G = g, \boldsymbol{A_k} = \boldsymbol{a}, k\right] f_{A_k}(\boldsymbol{A_k} = \boldsymbol{a} | T = 1, k) da$ is the standardized average potential outcome for group $G = g$ under intervention $\tilde{z}$, integrated over the distribution of allowable covariates $\boldsymbol{A_k}$ in the standard population $T = 1$, at time $k$.

On the additive scale, the intervention effect on disparity is defined as:

$$\tau_k(\tilde{z}) = \psi_k(\tilde{z}) - \psi_k(\tilde{z}_0) \quad (2)$$

which is the difference in disparity under intervention $\tilde{z}$ compared to the reference strategy $\tilde{z}_0$.

### 3.2 Enrollment windows

We define the enrollment window as a narrow span in calendar time, denoted by $k$. In medical settings, the length of the window (e.g., days, months) can reflect a clinical timeframe wherein treatment decisions of one window are unrelated to those of earlier windows.[12,16] Within each window, individuals are assessed for eligibility, and allowable covariates must be defined and measured by the end of the window. Persons may appear in multiple windows over time but may enroll only once per window. This structure conceptualizes a series of trials, each with a clearly defined target population at its corresponding calendar time and facilitates the following stratified sampling procedure (discussed shortly) that balances allowable covariates across groups at time $k$. It also makes temporal heterogeneity explicit and amenable to interpretable aggregation or modeling. Both disparities and the effects of interventions may vary over time due to evolving clinical protocols, fluctuating resource constraints, or broader contextual changes.[17] Indexing each trial to a calendar time $k$ allows investigators to examine, model, or aggregate results across the distribution of enrollment times, rather than implicitly assuming homogeneity over time (See Appendix for more details on the modeling and aggregation of calendar time).

In our motivating example, we used a 12-hour enrollment window under the conjecture that diagnostic work-ups induced by $SpO_2$ would be unrelated if separated by 12 or more hours. Conceptually, our trial specification assumes that treatment decisions are guided by $SpO_2$, with $SaO_2$ masked. In this idealized trial,[18] prior diagnostic work-ups would not influence subsequent decisions, even if the same individual were eligible in consecutive windows.

### 3.3 Enrollment groups

We define the social groups among whom disparities are measured based on characteristics thought to reflect systems of historical and/or contemporary (dis)advantage (see [19-22] for further discussion). Here, we focus on race and ethnicity, denoted by $G$, with the two groups of interest being NHB ($G = 1$) and NHW ($G = 0$). Our choice is motivated by the long-standing evidence of degraded PO accuracy for persons with dark skin tones.[23-26] While skin tone is unmeasured in EMR, it is strongly correlated with Black and White race,[27,28] and thus our choice leverages a meaningful proxy for differences in skin pigmentation to assess the population impacts of PO bias. In our example, we do not examine Hispanic or mixed racial/ethnic groups for two reasons: 1) our data used to emulate the TS+TT design cannot distinguish racial categories within the Hispanic category, where there is substantial heterogeneity in racial composition and thus skin tone, and 2) insufficient sample size.

### 3.4 Allowable covariates

We compare social groups who are similarly situated on "allowable" covariates, denoted by $\boldsymbol{A_k}$, those whose differential distribution is not considered to lead to disparity. The moral reasoning underlying allowability depends on the type of outcome being studied—specifically, whether the outcome represents a health care good (e.g., treatment receipt) or a health status. For treatment outcomes (e.g., dexamethasone), allowable covariates typically include indicators of clinical need that are considered as appropriate for guiding treatment decisions, consistent with established medical guidelines.[11,29] For health status outcomes (e.g., mortality), allowable covariates may include those that, at baseline, have a favorable distribution for the disadvantaged group, so as to avoid masking estimates of disparity[5] (although there are arguments against such adjustments for health status outcomes).[30,31] All other covariates used in the analysis are considered non-allowable, denoted by $\boldsymbol{N_k}$, and are included to account for potential biases such as confounding or selection bias but do not define the measure of disparity or the intervention effect. In our example, because the outcome is dexamethasone treatment, age, sex, comorbidities, vitals, laboratory measurements, and indicators of clinical severity (e.g., steroid and vasopressor use in the inpatient encounter but prior to $SpO_2$ measurement) reflect legitimate differences in clinical need within the current decision window and are therefore considered allowable. In contrast, smoking status and measures of socioeconomic status (e.g., the Area Deprivation Index (ADI)) are treated as non-allowable because they do not represent clinical need for treatment decisions and may reflect underlying social disadvantage or potentially lead to implicit bias in treatment decision-making. Since these factors may influence both oxygen saturation levels and treatment decisions,[13,32] we include them to adjust for potential confounding but not to define the disparity's measure.

### 3.5 Standard population

We define a within-study standard population, denoted by $T = 1$, to guide the balancing of allowable covariates across social groups during enrollment.[12] At each calendar time $k$, we sample persons so that the distribution of allowable covariates $\boldsymbol{A_k}$ is the same across social groups and matches that of the standard population. Under intervention effect heterogeneity (across allowable covariates), the choice of $T = 1$ can impact the magnitude and direction of the effect. Selecting the disadvantaged group helps to estimate excess potential benefits and harms for this population in medical and public health planning.[12] We choose the NHB group based on evidence of worse PO accuracy among Black patients.[23] Note that, with multiple social groups, it is possible to specify simultaneous TS+TT designs each with their own standard population.[12]

### 3.6 Enrollment process (Stratified sampling)

At each calendar time $k$, eligible persons enroll through a stratified sampling process conducted separately within each social group. First, we identify eligible persons from each group based on criteria measured during the enrollment window, mapping to a well-defined population at calendar time $k$. Second, within each group, we sample persons so that the distribution of allowable covariates $\boldsymbol{A_k}$ matches that of the within-study standard population $T = 1$. This sampling procedure ensures that any disparity in outcomes is not due to differences in allowable covariate distributions across groups. Although conceptual, a schematic example of how such stratified sampling could be implemented is provided in the Appendix.

### 3.7 Intervention strategies

Intervention strategies specify the actions or exposures whose effects the investigator seeks to evaluate. In applied settings, these strategies may compare two treatments or alternative policies. They may be deterministic or stochastic, and may assign categorical (e.g., medication type) or continuous exposures (e.g., levels of environmental pollution). Intervention strategies define the distribution of the exposure $Z$ that would be assigned in each arm of the trial. It is possible to devise strategies that depend only on covariates that are ethically appropriate for guiding the intervention's allocation of treatment. This raises the concept of "intervention-allowability"[5] that we address in the Discussion. In our example, we consider hypothetical scenarios in which PO devices exhibit different degrees of bias, producing different distributions of $SpO_2$ for individuals with darker versus lighter skin tones. Accordingly, we conceptualize each intervention strategy as a stochastic intervention that sets the $SpO_2$ for each social group to the distribution that would arise under a device with a specified level of bias.

Under the hypothetical trial, both providers and patients are assumed to be masked to $SaO_2$. For example, $SaO_2$ might be continuously monitored in a way that is not disclosed to clinical staff or patients, such that treatment decisions are guided solely by $SpO_2$.

### 3.8 Stratified (Within-group) randomization

After enrollment, we immediately randomize persons within each social group $G = g$ to intervention arms. This setup achieves two forms of balance: first, the stratified sampling procedure in the earlier step 3.6 ensures that allowable covariates are balanced across social groups, reflecting a meaningful disparity estimand. Second, the randomization procedure in this step balances both allowable covariates $\boldsymbol{A_k}$ and non-allowable covariates $\boldsymbol{N_k}$ across intervention arms within each group, ensuring exchangeability of intervention arms.

## 4. Emulation of the TS+TT using secondary data

### 4.1 Overview

Although we could, in theory, implement the TS+TT protocol to evaluate how PO bias influences disparities, doing so would be ethically and logistically infeasible. For example, our motivating example would require assigning black persons to devices with known unequal accuracy. Therefore, we instead emulate the TS+TT using secondary data, mapping each component (**Table 1**) to information available in EMR. To emulate components that are not directly observed, such as sampling-based enrollment, within-group randomization, and calendar time aggregation, we apply a sequential regression and prediction procedure, known as G-computation.[6,33] Beyond our example, other approaches including clone-censor-weight methods could be used for emulation using weights that emulate the sampling and randomization steps.[6,34]

### 4.2 Data structure

We construct a longitudinal dataset for emulating TS+TT. A column of calendar time $k$ defines the enrollment window. At each $k$, we identify eligible persons and record their social group membership $G$. Each person $i$ contributes one row per calendar time $k$ when eligibility criteria are met and may appear in multiple windows over time. We include a variable of exposure $Z$ (e.g., $SpO_2$). We select allowable covariates $\boldsymbol{A_k}$ (e.g., clinical need) to similarly situate social groups within each enrollment window and define a standard population $T = 1$ (e.g., NHB) that determines the distribution of $\boldsymbol{A_k}$ to which groups are standardized. We also include non-allowable covariates $\boldsymbol{N_k}$ (e.g., ADI) to control for confounding. Lastly, we attach the outcome $Y_{k+J}$ (e.g., dexamethasone treatment) for each record indexed by $k$, and the duration (denoted as $J_k$) from time zero to the time of either the outcome or administrative censoring, whichever occurs first.

In summary, each row of the analytic dataset is indexed by the vector $(i, k, G, T, Z, \boldsymbol{A_k}, \boldsymbol{N_k}, Y_{k+J}, J_k)$ for a person $i$ at time $k$. The data only include calendar times when all social groups of interest are represented, to ensure that aggregated disparity reflects an aggregation of estimates from well-defined populations across calendar time.

### 4.3 Identification and estimation

The counterfactual disparity (1) and intervention effect on disparity (2) are not directly observed. To identify these quantities from observational data, we rely on a set of identification assumptions: overlap and innocuous sampling[12] (TS component), and additionally conditional exchangeability, positivity, and consistency[35] (TT component).

We require overlap[12] in the allowables' distribution: At each calendar time $k$, persons of each social group $G = g$ are observed at each level of the allowable covariates $\boldsymbol{A_k}$ observed within the standard population ($T = 1$). Under innocuous sampling, the sampling procedure itself does not change how outcomes are conditionally distributed.[12] We assume that the effect of the intervention $\tilde{z}$ is unconfounded given the allowables $\boldsymbol{A_k}$ and non-allowable $\boldsymbol{N_k}$, separately within each social group $G$, at each time $k$. We further assume positivity and consistency.[35] Formal definitions of these assumption appear in the Appendix.

Under these assumptions, we can identify the aggregated standardized average potential outcome $\mu_k^{\tilde{z}}(g)$ as an iterated conditional expectation

$$\mu_k^{\tilde{z}}(g) = E_{\boldsymbol{A_k},k|T=1}\left[E_{Z,N_k|G=g,\boldsymbol{A_k},k}\left[E\left[Y_{k+J}\middle|G = g, Z, \boldsymbol{N_k}, \boldsymbol{A_k}, k\right]\middle|G = g, \boldsymbol{A_k}, k\right]|T = 1\right] \tag{3}$$

where $E_{Z,N_k|G=g,\boldsymbol{A_k},k}[\cdot]$ is an expectation over the group-specific joint distribution of $SpO_2$ values $Z$ under the intervention strategy $\tilde{z}$ and observed non-allowables $\boldsymbol{N_k}$, i.e., $f_{\tilde{z},\boldsymbol{N_k}}(Z, N_k|G = g, \boldsymbol{A_k}, k)$, and $E_{\boldsymbol{A_k},k|T=1}[\cdot]$ is over the distribution of allowables $\boldsymbol{A_k}$ and calendar time $k$ in the standard population $T = 1$, i.e., $f_{\boldsymbol{A_k},k}(\boldsymbol{A_k}, k|T = 1)$.

In this expression, counterfactual means for different social groups $G = 1$ and $G = 0$ under the same strategy, i.e., $\mu_k^{\tilde{z}}(1)$ and $\mu_k^{\tilde{z}}(0)$, are expectations over the same distribution of allowables $\boldsymbol{A_k}$ but potentially different distributions of non-allowables $\boldsymbol{N_k}$, so as to measure disparity. Whereas counterfactual means for each social group $G = g$ under different strategies, e.g., $\mu_k^{\tilde{z}_0}(g)$ and $\mu_k^{\tilde{z}_1}(g)$, are expectations over the same joint distribution of allowables and non-allowable confounders, so as to permit causal inference. Therefore, the identification formula adjusts for both allowables and non-allowables but handles them differently to permit both disparity measurement and causal inference.

We implement an adaptation of the G-computation procedure[6] to estimate (3). Here, we describe the steps for the binary outcome of treatment. In the Appendix we present the G-computation steps for time-to-treatment summarized using the restricted mean survival time (RMST).[36]

Step 1: Outcome modeling.

For each social group $G = g$, model the observed outcome $Y_{k+J}$ as a function of observed $SpO_2$ value $Z$, allowables $\boldsymbol{A_k}$, non-allowables $\boldsymbol{N_k}$, and calendar time $k$. For example, using a generalized linear model:

$$h\left(E\left[Y_{k+J}\middle|G = g, Z, \boldsymbol{A_k}, \boldsymbol{N_k}, k\right]\right) = \boldsymbol{X'\beta_g} \tag{4}$$

where $h(\cdot)$ is some link function (e.g., logistic for binary outcome), $\boldsymbol{X} = \{Z, \boldsymbol{A_k}, \boldsymbol{N_k}, k\}$ is the vector of the covariates and calendar time, and $\boldsymbol{\beta_g}$ is the group-specific coefficient vector. Note that $\boldsymbol{X}$ may also be replaced by a flexible function of the covariates (e.g., splines).

Step 2: Intervention assignment and prediction under the outcome model.

Assign all persons in group $G = g$ to receive strategy $\tilde{z}$. In our example, this involves simulating the $SpO_2$ distribution that would be produced by a device with the bias profile defined by $\tilde{z}$, i.e., $Z^* \sim f_{\tilde{z}}(Z|G = g, \boldsymbol{N_k}, \boldsymbol{A_k}, k)$. Starting from $SaO_2$, generate the corresponding $SpO_2$ by adding a random error drawn from the group-specific error distribution defined by the device's bias parameters. To reduce Monte Carlo variability, replicate the dataset $R$ times (e.g., $R = 10$) and draw an independent error term for each replicate. Then obtain predicted values using the group-specific outcome model (4) from Step 1 on this dataset where $SpO_2$ has been simulated, with $\boldsymbol{X} = \boldsymbol{X}^*$, where $\boldsymbol{X}^* = \{Z = Z^*, \boldsymbol{A_k}, \boldsymbol{N_k}, k\}$, and call these predictions $Q_g[1]$.

Step 3: Standardization model.

For each social group $G = g$, regress $Q_g[1]$ on the allowables $\boldsymbol{A_k}$ and calendar time $k$:

$$h\big(Q_g[1]\big|G = g, \boldsymbol{A_k}, k\big) = \boldsymbol{W}'\boldsymbol{\lambda_g} \tag{5}$$

where $\boldsymbol{W} = \{\boldsymbol{A_k}, k\}$ is the vector of allowable covariates (only) and calendar time and $\boldsymbol{\lambda_g}$ is the group-specific coefficient vector.

Step 4: Prediction under the standardization model.

Obtain predicted values using the group-specific model (5) from Step 3 on the empirical distribution of $\boldsymbol{A_k}$ and $k$ in the standard population $T = 1$, labeling them $Q_g[2]$.

Step 5: Marginalization among the standard population.

Take the average of $Q_g[2]$ among the standard population $T = 1$ to obtain $\mu_k^{\tilde{z}}(g)$, the aggregated standardized mean outcome for group $G = g$ under intervention $\tilde{z}$.

In summary, steps 1-2 estimate the conditional group-specific counterfactual mean under each intervention by averaging over the simulated distribution of $Z$ and $\boldsymbol{N_k}$ within levels of $\boldsymbol{A_k}$ and $k$ among $G = g$. Steps 3-5 then standardize these counterfactual outcomes over $\boldsymbol{A_k}$ (and calendar time $k$) among $T = 1$ to permit disparity estimation across groups under the same intervention strategy.

## 5. Data application

### 5.1 Methods

We applied our framework to EMR data from the Johns Hopkins COVID-19 Registry (JH-CROWN; March 2020–June 2024) to evaluate how differential accuracy in PO affect disparities in dexamethasone treatment among hospitalized COVID-19 patients. To characterize differential measurement error, we selected observations with paired $SpO_2$ and $SaO_2$ measurements recorded within 10 minutes of one another within each enrollment window.[15] To emulate our TS+TT design, we further restricted to instances where $SaO_2$ was measured after $SpO_2$. To focus on clinically relevant decision thresholds in our TS+TT emulation, analyses were restricted to $SaO_2$ values between 88% and 94%, a range in which measurement inaccuracy is most likely to affect treatment initiation and timing.[15] Race and ethnicity were self-reported and categorized according to the 2024 OMB standards,[37] using the "most frequent multiple responses" approach (See details in the Appendix). The study was approved by the Johns Hopkins School of Medicine Institutional Review Board.

We modeled nine hypothetical intervention arms (**Table 2**), corresponding to PO devices that differed in the degree of differential accuracy across racial groups. Each strategy specified group-specific distributions of measurement error that generated observed $SpO_2$ values from underlying $SaO_2$, with one reference strategy representing nondifferential accuracy. These strategies were designed to reflect empirically plausible ranges of measurement error observed in prior studies.[15,32,38-48]

Time zero was defined as the time of $SpO_2$ measurement, with follow-up continuing until dexamethasone treatment or administrative censoring at 24 hours. We evaluated the additive disparity and intervention effect in two outcomes: (1) dexamethasone receipt within 24 hours, and (2) time-to-dexamethasone treatment. To emulate the TS+TT, we followed the approach described in Section 4, with allowable and non-allowable covariates specified as shown in **Table 1**. To estimate uncertainty that account for clustering by person, we use a non-parametric, balanced, stratified cluster bootstrap that resamples persons with replacement.[6,49] We handled missing covariates using multiple imputation by chained equations, implemented via the *mice* package in R.[50] We also evaluated our model specification by comparing observed outcomes with those estimated under a natural course scenario.[51] To evaluate the robustness of our findings, we conducted two sensitivity analyses. The first analysis responds to the potential issue that clinical providers were unmasked to $SaO_2$ and thus disregarded the $SpO_2$ reading. As detailed in the Appendix, we swapped the roles of $SaO_2$ and $SpO_2$ in the G-computation procedure to simulate scenarios where providers respond to the device's $SpO_2$ as if it were the best available information for clinical decision-making. The second analysis excluded observations with $SaO_2$ measurements in the prior enrollment window to probe potential residual influence of prior diagnostic testing. All statistical analyses were performed in R version 4.5.2. Additional implementation details of the imputation, model evaluation, and sensitivity analyses appear in the Appendix.

### 5.2 Results

*Observed measurement error in pulse oximetry*

Among 15,122 $SpO_2$-$SaO_2$ pairs (7,350 from NHB; 7,772 from NHW) used to characterize differential measurement error, NHB patients had a higher mean (2.0% vs. 0.6%) and greater standard deviation (3.6% vs. 3.3%) (**sTable 1**). Occult hypoxemia ($SaO_2$ < 88% while $SpO_2$ ≥ 92%) was more common among NHB patients. The cohort for the TS+TT emulation included 1,883 eligible observations (907 from NHW; 976 from NHB) from 460 trials (**sTable 7**). Detailed cohort characteristics are provided in the Appendix (**sTable 2**).

*Observed additive disparities in 24-hour dexamethasone treatment receipt (%) and time-to-treatment*

Before standardizing on allowable covariates, NHB patients were 6.4% (95% CI: -13.4 to 0.6) less likely to receive treatment compared to NHW patients and experienced longer time to treatment (RMST difference: 54.2 minutes, 95% CI: -5.5 to 114.0) (**Table 3**). After standardization, the estimated additive disparities were attenuated: treatment receipt disparity was -2.2 % (95% CI: -8.2 to 3.7), and RMST disparity was 23.6 minutes (95% CI: -27.0 to 74.2). These results illustrate how the TS component clarifies differences that are *not* attributable to differences in allowable factors when measuring disparity.

*Simulated additive disparities in dexamethasone treatment and additive intervention effects*

Under the reference device with equal accuracy, the additive treatment disparity was -2.7% (95% CI: -8.8 to 3.3), and the RMST disparity was 30.0 minutes (95% CI: -26.3 to 86.3) (**Table 4**). Across simulated devices introducing varying degrees of differential measurement error, additive disparities were similar in magnitude to those under the reference scenario, with small intervention effects. For example, under the intervention with $\Delta\mu = 2$ and $\Delta\sigma = 0$, disparities in treatment receipt and RMST changed little from the reference, with estimated effects of 0.9% (95% CI: -0.9 to 2.8) and -1.1 minutes (95% CI: -6.9 to 4.8), respectively. Overall, disparities persisted across all simulated scenarios and were only slightly altered across forms of differential measurement error.

*Sensitivity analyses*

In the sensitivity analysis that swapped $SpO_2$ and $SaO_2$, simulated disparities were generally larger and increased with the severity of differential accuracy (**sTable 3 vs. Table 4**). For example, the intervention with $\Delta\mu = 2$ and $\Delta\sigma = 3$—the most biased device—exacerbated the treatment disparity than the reference (-10.7% vs. -5.3%), corresponding to an intervention effect of -5.4% (95% CI: -7.9 to -2.8). Results were similar to the primary analysis when excluding observations with $SaO_2$ measurements in the prior window (**sTable 4 vs. Table 4**).

## 6. Discussion

We have proposed a novel TS+TT framework for evaluating intervention effects on disparities. The TS component grounds disparity measurement in allowability-based ethical principles and anchors it to an explicit population in calendar time through stratified sampling, while the TT component conceptualizes randomization of interventions within social groups. Nesting TS within TT yields well-defined causal estimands for how interventions modify disparities and provides a transparent, general framework to inform policy and clinical decision-making centered on fairness in health and healthcare.

In our application, we defined "outcome-allowable" covariates appropriate for defining the outcome disparity, but our TS+TT framework is flexible enough to accommodate intervention-allowable covariates. These covariates reflect normative judgments about which characteristics are considered morally appropriate for allocating exposure across social groups[5] and, in this sense, help define the interventions. For instance, one could define interventions that assign different degrees of permissible bias in $SpO_2$ within levels of $SaO_2$ (e.g., 70-85%, ≥85%). In fact, recent FDA draft guidance specifies $SaO_2$-stratified performance thresholds that permit larger bias at lower $SaO_2$ levels (e.g., 3.5% for $SaO_2$ 70-85%) and smaller bias at higher $SaO_2$ levels (e.g., 1.5% for $SaO_2$ ≥85%). This implicitly treats $SaO_2$ as an allowable variable that governs how much bias in device performance is considered tolerable, likely reflecting engineering constraints. We do not take a position on specific allowability choices in this application and do not implement this approach due to limited sample size within $SaO_2$ strata. However, the framework is explicitly structured to support such extensions when allowability criteria are clearly specified and data permit. Specifying strategies that assign treatment according to intervention-allowable covariates is motivated for strategies intended to be salutary.

Our framework provides a practical tool for policy and regulatory decision-making. In the context of PO bias, it enables evaluation of how varying degrees of bias affect treatment disparities and can inform evidence-based standards for device performance. More broadly, it can be applied to other measurement or decision systems—such as predictive algorithms or clinical rules[52-56]—to assess how algorithmic design choices shape downstream disparities and to inform interventions that reduce, rather than inadvertently exacerbate, health disparities.

Our data application faced limitations inherent to emulating an idealized trial with real-world data. First, masking clinicians to $SaO_2$ was infeasible in routine clinical care. Although we approximated masking, residual awareness or unmeasured clinical need may still exist. In addition, our application estimated an intention-to-treat effect of baseline biased $SpO_2$ but did not fully capture the cumulative impact of sustained measurement bias over the duration of care (i.e., subsequent $SpO_2$ readings during follow-up). Turning to the magnitude of the disparities, the risk of adverse consequences from estimated delays in treatment (approximately 30 minutes across scenarios) among NHB patients are unclear. We suspect that such risks are low for any individual but may be noticeable at the population-level,[57] which could be investigated by examining the impacts of bias thresholds on downstream adverse outcomes. Weak intervention effects may also reflect features of this specific clinical context, such as standardized treatment protocols and the use of multiple clinical indicators beyond $SpO_2$.[58] In light of that, we conducted analyses restricted to patients near treatment-relevant clinical thresholds ($SaO_2$ 88-94%), where inaccurate $SpO_2$ readings are most likely to influence treatment decisions.[15] Our sensitivity analyses that swapped $SaO_2$ and $SpO_2$ yielded larger disparities and intervention effects, providing insight into settings where PO bias may be more consequential when treatment decisions rely more heavily on $SpO_2$. Furthermore, estimation relied on standard causal assumptions. While most of these assumptions are either untestable or difficult to fully verify with EMR data, we have discussed their implications and potential challenges in detail in the Appendix. Finally, our data pertained to a single regional health system. We note that the TS component can specify sampling strategies that can be emulated to support generalizability or transportability to well-defined populations of interest,[12] which can be adapted to TS+TT.

As discussed in the Appendix, TS+TT shares much in common with other approaches involving hypothetical interventions in the disparities setting, such as causal decompositions that account for allowability,[5] or approximations of causal mediation estimands that intervene on group membership.[59] But the TS+TT framework is more universal. It allows multiple intervention strategies, aligns naturally with trial-based policy questions, and extends naturally to longitudinal settings with dynamic treatment strategies. In this sense, causal decompositions are a special case that TS+TT can subsume, though TS+TT is not itself a decomposition nor mediation estimand.

**Table 1. TS+TT Protocol Specification and Emulation**

| | Specification | | Emulation |
|---|---|---|---|
| | Enrollment windows | 12-hour windows from March 2020 to June 2024 | Same, constructed from EMR timestamps |
| | Enrollment groups | Self-reported NHB and NHW | Mapped to self-reported race and ethnicity |
| | Eligibility criteria | Adults aged ≥18 years receiving care in the Johns Hopkins Health System, with confirmed COVID-19 diagnosis, with $SpO_2$ and $SaO_2$ measured concurrently | Mapped to first eligible paired $SpO_2$-$SaO_2$ measurement within each 12-hour enrollment window, with $SaO_2$ measured after $SpO_2$ |
| | Allowable covariates | $SaO_2$, age, sex, BMI, pulse rate, respiratory rate, body temperature, systolic blood pressure, comorbidities (Elixhauser comorbidity index), bilirubin, hemoglobin, creatinine, prior steroid use, vasopressor use prior to $SpO_2$ measurement | Mapped to corresponding EMR data fields and timestamps |
| | Standard population | NHB | Same |
| Causal estimand | Enrollment process | Stratified sampling so that the distribution of allowable covariates of each social group matches that of the within-study standard population | Statistical implementation via G-computation (Steps 3-5) to balance allowable covariates across social groups |
| | Intervention strategies | Hypothetical PO devices with prespecified race-specific measurement error distributions[a] (arm definitions in **Table 2)**<br><br>Providers and patients are masked to the assigned device and $SaO_2$ values (e.g., $SaO_2$ might be continuously monitored in a way that is not disclosed to clinical staff or patients) | Simulate $SpO_2$ by adding random error drawn from prespecified race-specific error distributions to $SaO_2$<br><br>Masking of $SaO_2$ is approximated by restricting to observations in which $SaO_2$ was measured after the corresponding $SpO_2$ |
| | Assignment | Stratified randomization, in which devices are randomly assigned to individuals within each social group to balance confounders (allowable and non- | Statistical implementation via G-computation (Steps 1-2) to balance both allowable and non-allowable covariates across |
| | | allowable covariates) across intervention arms | intervention arms within each social group. Non-allowable covariates included ADI and smoking status |
| | Outcome | 24-hour dexamethasone treatment (% treated and time-to-treatment) | Mapped to dexamethasone medication administration timestamps |
| | Follow-up | Starts at time of $SpO_2$ measurement; ends at dexamethasone treatment or 24 hours, whichever occurs first | Same |
| | Causal contrast | Intention-to-treat (i.e., effect of assignment to each hypothetical device) | Same |
| Identifying assumptions | Overlap, innocuous sampling, randomized assignment | | Overlap, innocuous sampling, conditional exchangeability, positivity, and consistency assumptions involving both allowable and non-allowable covariates (Details in Appendix**)** |
| Statistical estimand and assumptions | Additive disparity and additive intervention effect on disparity, for both % treated and RMST under the intention-to-treat principle, aggregated over calendar time<br><br>Assumption of missing-at-random for imputation<br>Multiple imputation for missing data<br>Assumption of correct model specification | | Same. Assumption of correct model specification evaluated by modeling natural course<br><br>Same<br><br>Sensitivity analyses of swapping $SpO_2$ and $SaO_2$ and excluding observations with $SaO_2$ measurements in the prior enrollment window |

Abbreviations: EMR = Electronic Medical Records, NHW = Non-Hispanic White, NHB = Non-Hispanic Black, PO = Pulse Oximeter, BMI = Body Mass Index, RMST = Restricted Mean Survival Time, ADI = Area Deprivation Index

[a] Measurement error was defined as the difference between the $SpO_2$ and the $SaO_2$ ($\varepsilon = SpO_2 - SaO_2$), with the mean error denoted by $\mu$ and standard deviation by $\sigma$. Racial bias was summarized by $\Delta\mu$ and $\Delta\sigma$, the difference in $\mu$ and $\sigma$ of $\varepsilon$ in the NHB ($G = 1$) versus NHW ($G = 0$) group.

**Table 2. Hypothetical PO intervention arms defined by group-specific distributions with bias parameters**

| Intervention arm | G = 0 | G = 1 | Δμ | Δσ |
|---|---|---|---|---|
| $\tilde{z}_0$ (Ref) | N(0.62, $1.61^2$) | N(0.62, $1.61^2$) | 0 | 0 |
| $\tilde{z}_1$ | N(0.62, $1.61^2$) | N(2.62, $1.61^2$) | 2 | 0 |
| $\tilde{z}_2$ | N(0.62, $1.61^2$) | N(1.62, $1.61^2$) | 1 | 0 |
| $\tilde{z}_3$ | N(0.62, $1.61^2$) | N(0.62, $4.61^2$) | 0 | 3 |
| $\tilde{z}_4$ | N(0.62, $1.61^2$) | N(0.62, $3.11^2$) | 0 | 1.5 |
| $\tilde{z}_5$ | N(0.62, $1.61^2$) | N(2.62, $4.61^2$) | 2 | 3 |
| $\tilde{z}_6$ | N(0.62, $1.61^2$) | N(2.62, $3.11^2$) | 2 | 1.5 |
| $\tilde{z}_7$ | N(0.62, $1.61^2$) | N(1.62, $4.61^2$) | 1 | 3 |
| $\tilde{z}_8$ | N(0.62, $1.61^2$) | N(1.62, $3.11^2$) | 1 | 1.5 |

Note: N(μ, $\sigma^2$) denotes a Normal distribution with mean μ and variance $\sigma^2$. $\tilde{z}_0$ served as the reference device, with Δμ = 0 and Δσ = 0, indicating no racial bias in either mean error or variability. $\tilde{z}_1$ (or $\tilde{z}_2$) introduced higher mean error in the NHB group, Δμ = 2 (or 1), with equal variability, Δσ = 0. $\tilde{z}_3$ (or $\tilde{z}_4$) maintained equal mean error, Δμ = 0, but increased variability in the NHB group, Δσ = 3 (or 1.5). $\tilde{z}_5$, $\tilde{z}_6$, $\tilde{z}_7$, $\tilde{z}_8$ combined both elevated mean error and variability in the NHB group (Δμ = 1 or 2, Δσ = 1.5 or 3).

**Table 3. Observed additive disparities in 24-hour dexamethasone treatment receipt (%) and time-to-treatment**

| | NHW (N = 907) | NHB (N = 976) | Additive disparity |
|---|---|---|---|
| % Treated | | | |
| Before standardization | 26.7 (21.3, 32.1) | 20.3 (15.8, 24.8) | -6.4 (-13.4, 0.6) |
| After standardization | 22.5 (16.7, 28.3) | 20.3 (15.8, 24.8) | -2.2 (-8.2, 3.7) |
| RMST, minutes | | | |
| Before standardization | 1231.2 (1188.2, 1274.3) | 1285.5 (1245.7, 1325.3) | 54.2 (-5.5, 114.0) |
| After standardization | 1269.9 (1220.4, 1319.4) | 1293.5 (1254.9, 1332.2) | 23.6 (-27.0, 74.2) |

Abbreviations: NHW = Non-Hispanic White, NHB = Non-Hispanic Black, RMST = Restricted Mean Survival Time

Note: Values are presented as point estimates, with 95% confidence intervals shown in parentheses.

**Table 4. Simulated additive disparities in dexamethasone treatment and additive intervention effects under varying degrees of racial bias**

| | NHW (N = 907) | NHB (N = 976) | Additive disparity | Additive intervention effect |
|---|---|---|---|---|
| **% Treated** | | | | |
| Reference ($\Delta\mu = 0$, $\Delta\sigma = 0$) | 21.8 (16.0, 27.6) | 19.1 (14.3, 23.8) | -2.7 (-8.8, 3.3) | Reference |
| Intervention | | | | |
| $\Delta\mu = 2$, $\Delta\sigma = 0$ | 21.8 (16.0, 27.6) | 20.0 (15.3, 24.6) | -1.8 (-7.5, 3.9) | 0.9 (-0.9, 2.8) |
| $\Delta\mu = 1$, $\Delta\sigma = 0$ | 21.8 (16.0, 27.6) | 19.5 (14.8, 24.1) | -2.3 (-8.1, 3.5) | 0.4 (-0.6, 1.4) |
| $\Delta\mu = 0$, $\Delta\sigma = 3$ | 21.8 (16.0, 27.6) | 19.4 (14.2, 24.7) | -2.4 (-9.0, 4.3) | 0.4 (-2.2, 3.0) |
| $\Delta\mu = 0$, $\Delta\sigma = 1.5$ | 21.8 (16.0, 27.6) | 19.2 (14.4, 24.1) | -2.6 (-8.7, 3.6) | 0.2 (-0.9, 1.3) |
| $\Delta\mu = 2$, $\Delta\sigma = 3$ | 21.8 (16.0, 27.6) | 20.2 (15.5, 24.9) | -1.6 (-7.5, 4.4) | 1.2 (-1.0, 3.3) |
| $\Delta\mu = 2$, $\Delta\sigma = 1.5$ | 21.8 (16.0, 27.6) | 20.1 (15.6, 24.6) | -1.7 (-7.4, 4.0) | 1.1 (-0.6, 2.7) |
| $\Delta\mu = 1$, $\Delta\sigma = 3$ | 21.8 (16.0, 27.6) | 19.8 (15.0, 24.7) | -2.0 (-8.2, 4.2) | 0.8 (-1.3, 2.9) |
| $\Delta\mu = 1$, $\Delta\sigma = 1.5$ | 21.8 (16.0, 27.6) | 19.7 (15.1, 24.2) | -2.1 (-8.0, 3.7) | 0.6 (-0.4, 1.6) |
| **RMST, minutes** | | | | |
| Reference ($\Delta\mu = 0$, $\Delta\sigma = 0$) | 1268.5 (1219.1, 1318.0) | 1298.6 (1255.6, 1341.5) | 30.0 (-26.3, 86.3) | Reference |
| Intervention | | | | |
| $\Delta\mu = 2$, $\Delta\sigma = 0$ | 1268.5 (1219.1, 1318.0) | 1297.5 (1257.5, 1337.5) | 28.9 (-24.7, 82.6) | -1.1 (-6.9, 4.8) |
| $\Delta\mu = 1$, $\Delta\sigma = 0$ | 1268.5 (1219.1, 1318.0) | 1298.0 (1256.5, 1339.4) | 29.4 (-25.5, 84.4) | -0.6 (-3.5, 2.4) |
| $\Delta\mu = 0$, $\Delta\sigma = 3$ | 1268.5 (1219.1, 1318.0) | 1295.8 (1258.8, 1332.8) | 27.3 (-23.6, 78.1) | -2.7 (-18.6, 13.2) |
| $\Delta\mu = 0$, $\Delta\sigma = 1.5$ | 1268.5 (1219.1, 1318.0) | 1296.8 (1258.7, 1334.9) | 28.2 (-23.6, 80.1) | -1.8 (-12.5, 8.9) |
| $\Delta\mu = 2$, $\Delta\sigma = 3$ | 1268.5 (1219.1, 1318.0) | 1295.3 (1258.1, 1332.4) | 26.8 (-24.1, 77.6) | -3.3 (-23.8, 17.3) |
| $\Delta\mu = 2$, $\Delta\sigma = 1.5$ | 1268.5 (1219.1, 1318.0) | 1296.2 (1258.9, 1333.5) | 27.7 (-23.5, 78.8) | -2.4 (-17.2, 12.5) |
| $\Delta\mu = 1$, $\Delta\sigma = 3$ | 1268.5 (1219.1, 1318.0) | 1295.6 (1258.6, 1332.6) | 27.0 (-23.8, 77.8) | -3.0 (-21.0, 15.1) |
| $\Delta\mu = 1$, $\Delta\sigma = 1.5$ | 1268.5 (1219.1, 1318.0) | 1296.5 (1258.9, 1334.1) | 28.0 (-23.4, 79.3) | -2.1 (-14.6, 10.5) |

Abbreviations: NHW = Non-Hispanic White, NHB = Non-Hispanic Black, RMST = Restricted Mean Survival Time

Note: Values are presented as point estimates, with 95% confidence intervals shown in parentheses.

## Appendix

### Table of Contents

**Supplemental Details on Data Application**

*Methods*

Race and ethnicity categorization

Electronic medical record (EMR) data from the Johns Hopkins Health System capture race and ethnicity using two variables: race and Hispanic ethnicity, consistent with the 2011 U.S. Department of Health and Human Services (HHS) implementation guidance.[1]

The 2024 U.S. Office of Management and Budget (OMB) Statistical Policy Directive No. 15 recommends collecting race and ethnicity using a combined question format with a minimum of seven categories.[2] The guidance also outlines multiple approaches for reporting these data, including: (1) the *alone or in combination* approach, (2) the *most frequent multiple responses* approach, and (3) the *combined multiracial and/or multiethnic* approach.

In line with OMB guidance, we adopted a "most frequent multiple responses" approach that reports as many possible race and ethnicity combinations as possible. Specifically, we constructed mutually exclusive categories based on combinations of race and Hispanic or Latino ethnicity.

We defined the following categories:

1. **Single race or ethnicity categories:**

- American Indian or Alaska Native
- Asian
- Black or African American
- Hispanic or Latino
- Native Hawaiian or Pacific Islander
- White (including Middle Eastern or North African)

2. **Race and ethnicity combination categories (Hispanic or Latino ethnicity with reported race):**

- American Indian or Alaska Native + Hispanic or Latino
- Asian + Hispanic or Latino
- Black or African American + Hispanic or Latino
- Native Hawaiian or Pacific Islander + Hispanic or Latino
- White (including Middle Eastern or North African) + Hispanic or Latino

Although we implemented the updated OMB standards as described above, our analysis in this example focused on non-Hispanic Black and non-Hispanic White groups. We did not examine Hispanic or mixed racial/ethnic groups for two reasons: (1) the data used to emulate the TS+TT design could not distinguish racial categories within the Hispanic category, which exhibits substantial heterogeneity in racial composition and thus skin tone; and (2) insufficient sample sizes for these groups.

Definitions of intervention arms

We modeled nine hypothetical intervention arms $\tilde{Z} \in (\tilde{z}_0, \tilde{z}_1, \tilde{z}_2, \tilde{z}_3, \tilde{z}_4, \tilde{z}_5, \tilde{z}_6, \tilde{z}_7, \tilde{z}_8)$ (**Table 2**), where each arm represents a PO device with varying levels of racial bias. Each device defines a bivariate distribution of measurement error across racial groups, based on parameters for the group-specific mean $\mu_g$ and standard deviation $\sigma_g$ for $g \in (1,0)$. For group $G = 0$ (NHW), we fixed the measurement error as $\varepsilon_0 \sim N(\mu_0, \sigma_0^2)$ with $\mu_0 = 0.62$ and $\sigma_0 = 1.61$, based on pooled estimates from the most comprehensive and up-to-date systematic review of PO measurement error.[3] For group $G = 1$ (NHB), each device specified a distinct distribution $\varepsilon_1 \sim N(\mu_1, \sigma_1^2)$, characterized by bias parameters $\Delta\mu = \mu_1 - \mu_0$ and $\Delta\sigma = \sigma_1 - \sigma_0$. To reflect realistic scenarios, we anchored these bias parameter values using the median value observed in an empirical review of published studies on $SpO_2$ measurement error among Black and White patients.[4-16] We then varied $\Delta\mu$ and $\Delta\sigma$ to three levels: the observed median, half of this observed median, and zero. Specifically, $\Delta\mu \in (2,1,0)$, $\Delta\sigma \in (3,1.5,0)$, resulting in nine distinct intervention arms with $\Delta\mu = \Delta\sigma = 0$ serving as the reference arm $\tilde{z}_0$. For example, under the device $\tilde{z}_1$, the NHW group was assigned $\mu_0 = 0.62$ and $\sigma_0 = 1.61$, while the NHB group was assigned $\mu_1 = 2.62$ and $\sigma_1 = 1.61$. To simulate each device, we drew $\varepsilon \sim N(\mu_g, \sigma_g^2)$ from the group-specific distribution and generated $SpO_2$ by adding this error to the $SaO_2$.

Sensitivity analyses

To evaluate the robustness of our findings, we conducted two sensitivity analyses. The first analysis responds to the potential issue that clinical providers were unmasked to $SaO_2$ and thus disregarded the $SpO_2$ reading. To address this, we implemented an alternative simulation approach by swapping the roles of $SpO_2$ and $SaO_2$ in the G-computation procedure. Specifically, we left step 1 (outcome modeling) unchanged, but in step 2 (intervention assignment) recoded the observed $SaO_2$ as $SpO_2$, and recoded the simulated $SpO_2$ as $SaO_2$, and proceeded with step 3 (standardized model) and step 4 (prediction) as described. This approach approximates a counterfactual scenario in which providers respond to simulated oxygen saturation as if it were their best available information, offering an upper bound on the potential impact of varying device bias. The second analysis excluded observations with $SaO_2$ measurements in the prior enrollment window to probe potential residual influence of prior diagnostic testing. All statistical analyses were performed in R version 4.5.2.

*Results*

The analytic cohort included 1,883 eligible observations from 460 trials (**sTable 7**), with 907 (48.2%) from NHW and 976 (51.8%) from NHB participants (**sFigure 4**). We reported characteristics from the first imputed complete dataset (**sTable 2**), as covariate distributions were similar across all imputed datasets. NHB patients were slightly younger on average (mean age 60.2 vs. 62.2 years) and had a lower proportion of males (53.0% vs. 59.1%) compared with NHW patients. Several clinical characteristics also differed between groups. NHB patients had higher pulse and respiratory rates, lower hemoglobin, higher creatinine, and lower total bilirubin. Prior steroid use was more common in NHW patients (79.6% vs. 67.9%). Socioeconomic status also differed substantially between groups: NHB patients resided in neighborhoods with greater deprivation, as reflected by higher mean ADI national percentiles (53.6 vs. 38.1, SMD = 0.61) and a greater concentration in higher ADI deciles. Other characteristics, including BMI, systolic blood pressure, overall comorbidity burden, vasopressor use at $SpO_2$ time, and smoking status, were similar across groups.

## Supplementary Figures

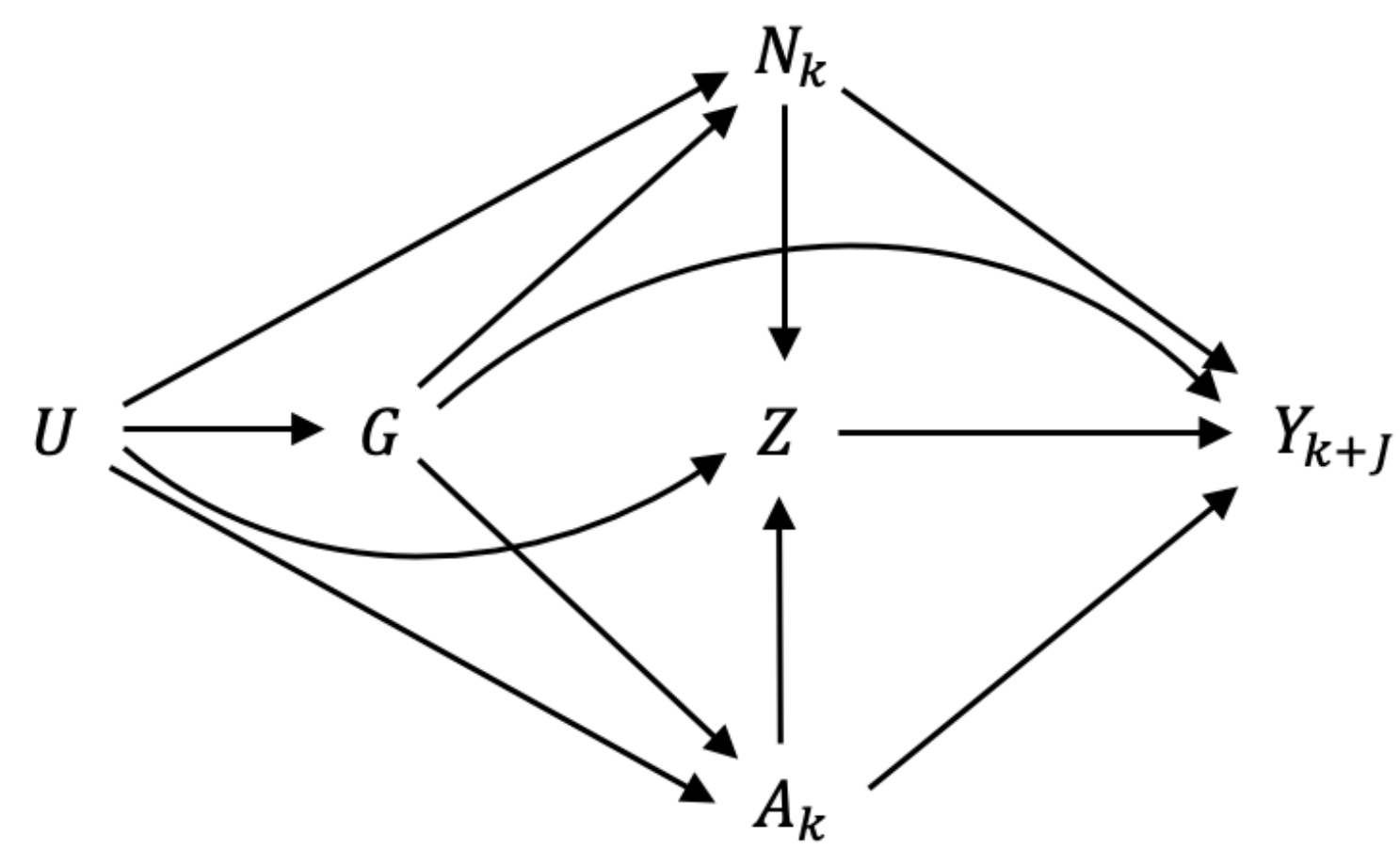

**sFigure 1. Causal directed acyclic graph (DAG) depicting the relationships between $\boldsymbol{Z}$ (i.e., $\boldsymbol{SpO_2}$), race and ethnicity $\boldsymbol{G}$, allowable covariates $\boldsymbol{A_k}$, non-allowable covariates $\boldsymbol{N_k}$, 24-hour dexamethasone treatment $\boldsymbol{Y_{k+J}}$, and historical processes $\boldsymbol{U}$**

The exposure $Z$ reflects device-measured oxygen saturation, which may be affected by racial bias and directly influences the treatment decision $Y_{k+J}$. Historical processes $U$ (e.g., structural forms of injustice[17] possibly involving paternal characteristics such as socioeconomic status, wealth, country of origin, or physical phenotypes) shape $G$, influence covariates ($\boldsymbol{A_k}$, $\boldsymbol{N_k}$) through pathways of discrimination and disadvantage,[18] and may directly affect $Z$ by driving inequities in device development and validation. $G \rightarrow \boldsymbol{A_k}$ reflects embodiment of social disadvantage,[19] where racialized social position affects clinical needs. $G \rightarrow (\boldsymbol{A_k}, \boldsymbol{N_k})$ also captures mechanisms of discrimination and internalized racism.[18] These covariates affect both $Z$ and $Y_{k+J}$ and serve as confounders of the $Z$-$Y_{k+J}$ relationship. We adjust for both $\boldsymbol{A_k}$ and $\boldsymbol{N_k}$ to control confounding across intervention arms within each social group. Only $\boldsymbol{A_k}$ are used to standardize the disparity measure within an intervention arm.

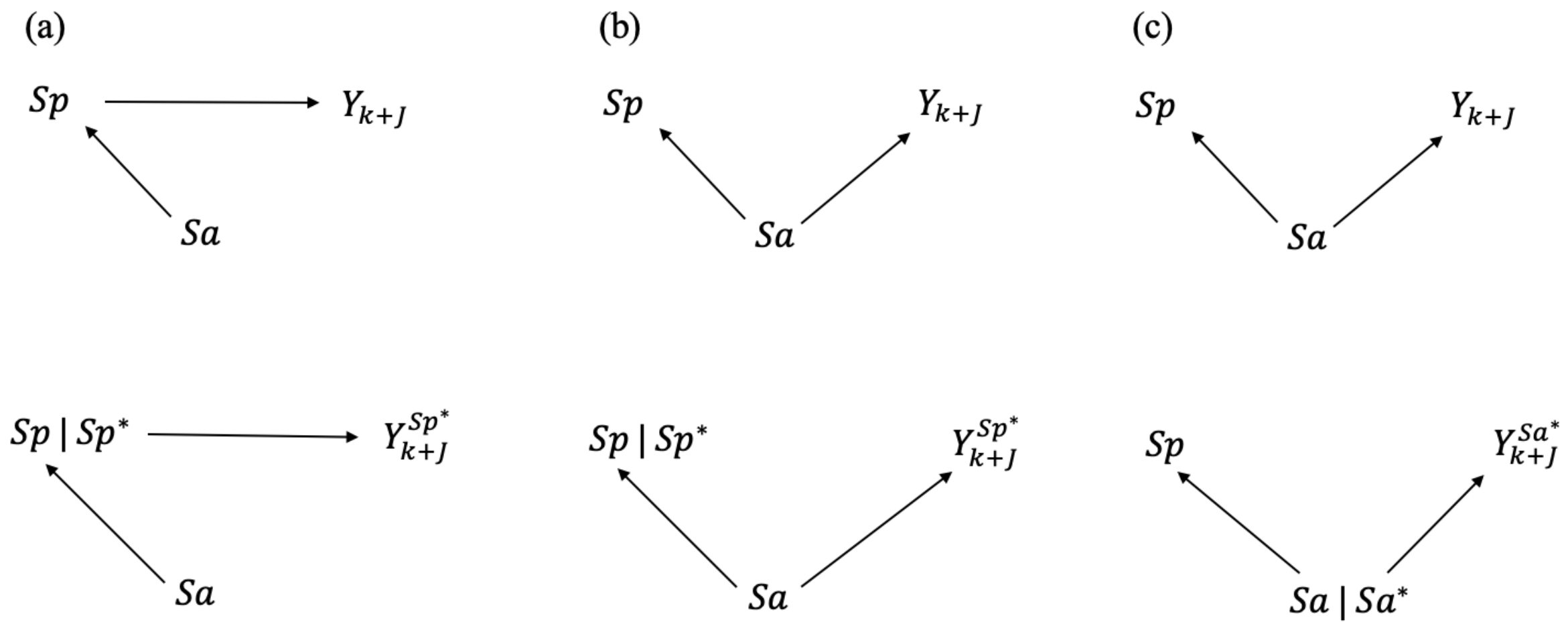

**sFigure 2. Causal diagrams (directed acyclic graphs [DAGs, top row] and single world intervention graphs [SWIGs, bottom row]) depicting three scenarios regarding provider knowledge of oxygen saturation**

(a) $SaO_2$ unknown (ideal target trial) (b) $SaO_2$ known (real-world setting and emulation) (c) $SaO_2$ known (sensitivity analysis)
Abbreviations: Sp = $SpO_2$, Sa = $SaO_2$, Sp* = intervened $SpO_2$, Sa* = intervened $SaO_2$, $Y_{k+J}$ = dexamethasone treatment, superscript indicates potential outcome.

Ideally, as depicted in panel (a), our target trial assumes that treatment decisions are, initially at least, made solely based on $SpO_2$, with $SaO_2$ unknown to providers. However, our primary emulation (panel b) more closely reflects real-world practice, where $SaO_2$ often becomes available to providers after the $SpO_2$ measurement and may inform or even drive treatment decisions. Although we attempted to mitigate this by restricting to cases where $SaO_2$ was measured after $SpO_2$, the potential for residual awareness remains. We observed minimal variation in treatment disparities across PO devices with differing levels of racial bias. This suggests that access to $SaO_2$ may have diminished the influence of $SpO_2$ on provider decision-making, such that intervening on $SpO_2$ may have limited effect on disparities.

To explore this possibility further, we conducted a sensitivity analysis (panel c) that reversed the roles of $SpO_2$ and $SaO_2$ in our G-computation procedure. Specifically, we implemented the following modifications:
Step 1: unchanged.
Step 2 (Intervention assignment): recoded $SaO_2$ as $SpO_2$, simulated $SpO_2$ based on the device-specific error model (and the original $SaO_2$), and then recoded the simulated values as $SaO_2$.
Step 3-5: unchanged.
This approach mimics a counterfactual scenario in which providers rely on simulated $SpO_2$ as their best available information, i.e., treating it as $SaO_2$. It allows us to estimate the potential upper bound of treatment disparities that could arise if decisions were based solely on $SpO_2$.

In the sensitivity analysis that swapped the roles of $SaO_2$ and $SpO_2$, simulated disparities were generally larger across intervention arms than in the primary analysis (**sTable 3 vs. Table 4**). Devices with higher mean error, greater variability, or both in NHB compared with NHW generated progressively larger disparities in treatment receipt than the reference device. Devices with higher mean error in NHB also generated larger disparities in RMST, although in other scenarios intervention effects remained small. Overall, the magnitude of intervention effects tended to increase with the severity of device bias, suggesting that when treatment decisions rely more heavily on $SpO_2$, racial bias in pulse oximeter accuracy may play a larger role in generating treatment disparities. Notably, this study focuses on hospitalized COVID-19 patients, a context in which clinical treatment protocols such as the widespread administration of dexamethasone may have constrained provider discretion.[20] These protocols may constrain provider discretion and reduce the extent to which $SpO_2$ directly informs treatment decisions, which may limit the generalizability of our findings to other settings.

Source population
NHB and NHW adults receiving care in the Johns Hopkins Health System with confirmed COVID-19 between March 2020 and June 2024, with at least one paired $SpO_2$-$SaO_2$ measurement within 10 minutes

| | NHB | NHW |
|---|---|---|
| Observations | 17791 | 18975 |
| Persons | 555 | 736 |
| Obs./person | 1 to 520 | 1 to 641 |

↓

Analytic population
Closest paired $SpO_2$-$SaO_2$ measurement within each 12-hour enrollment window

| | NHB | NHW |
|---|---|---|
| Observations | 7350 | 7772 |
| Persons | 553 | 732 |
| Obs./person | 1 to 186 | 1 to 155 |

**sFigure 3. Flowchart depicting construction of the analytic population for characterizing observed measurement error**

Source population
NHB and NHW adults receiving care in the Johns Hopkins Health System with confirmed COVID-19 between March 2020 and June 2024, with at least one paired $SpO_2$-$SaO_2$ measurement within 10 minutes

| | NHB | NHW |
|---|---|---|
| Observations | 17791 | 18975 |
| Persons | 555 | 736 |
| Obs./person | 1 to 520 | 1 to 641 |

Enrolled eligible population
First eligible paired $SpO_2$-$SaO_2$ measurement per trial, where each trial is indexed by a 12-hour calendar-time enrollment window, with $SaO_2$ measured after $SpO_2$

| | NHB | NHW |
|---|---|---|
| Observations | 5041 | 5409 |
| Persons | 482 | 607 |
| Obs./person | 1 to 133 | 1 to 145 |
| Trials | 2709 | |

Analytic population
$SaO_2$ values between 88% and 94%, trials with concurrent representation of both NHB and NHW groups

| | NHB | NHW |
|---|---|---|
| Observations | 976 | 907 |
| Persons | 232 | 232 |
| Obs./person | 1 to 45 | 1 to 47 |
| Trials | 460 | |

**sFigure 4. Flowchart depicting the construction of the analytic population for emulating the TS+TT design**

## Supplementary Tables

**sTable 1. Observed measurement error in pulse oximetry and occult hypoxemia by race/ethnicity**

| Race/ethnicity | No. of pairs | No. of unique patients | $SpO_2$, % | | $SaO_2$, % | | Measurement error ($SpO_2$-$SaO_2$), % | | Proportion of occult hypoxemia, % |
|---|---|---|---|---|---|---|---|---|---|
| | | | Median (IQR) | Mean (SD) | Median (IQR) | Mean (SD) | Median (IQR) | Mean (SD) | |
| NHW | 7,772 | 732 | 96.0 (5.0) | 95.4 (4.0) | 95.7 (4.7) | 94.8 (4.0) | 0.5 (3.2) | 0.6 (3.3) | 2.3 |
| NHB | 7,350 | 553 | 97.0 (4.0) | 96.3 (3.6) | 95.3 (5.4) | 94.4 (4.3) | 1.6 (3.4) | 2.0 (3.6) | 4.8 |
| P for difference | | | | <0.001 | | <0.001 | | <0.001 | <0.001 |

Abbreviations: NHW = Non-Hispanic White, NHB = Non-Hispanic Black, No. = Number, SD = Standard Deviation; IQR = Interquartile Range

**sTable 2. Cohort characteristics by race/ethnicity**

| | NHW (N = 907) | NHB (N = 976) | SMD |
|---|---|---|---|
| No. of unique patients | 232 | 232 | |
| Hospital location | | | **0.32** |
| Johns Hopkins Hospital | 557 (61.4) | 707 (72.4) | |
| Bayview Medical Center | 196 (21.6) | 116 (11.9) | |
| Howard County General Hospital | 116 (12.8) | 96 (9.8) | |
| Suburban Hospital | 28 (3.1) | 30 (3.1) | |
| Sibley Memorial Hospital | ≤11 (≤1.2) | >20 (>2.0) | |
| Unknown | ≤11 (≤1.2) | ≤11 (≤1.1) | |
| Age | 62.2 ± 14.6 | 60.2 ± 12.3 | **0.15** |
| Sex (Male %) | 536 (59.1) | 517 (53.0) | **0.12** |
| BMI, $kg/m^2$ | 33.5 ± 9.1 | 34.0 ± 9.8 | 0.06 |
| BMI category | | | 0.07 |
| Underweight (<18.5) | ≤11 (≤1.2) | ≤11 (≤1.1) | |
| Normal or overweight (18.5 to <30) | >360 (>39.7) | >360 (>36.9) | |
| Obesity (≥30) | 541 (59.6) | 609 (62.4) | |
| Pulse rate | 114.4 ± 23.9 | 117.1 ± 21.6 | **0.12** |
| Respiratory rate | 34.6 ± 8.9 | 37.0 ± 9.1 | **0.27** |
| Body temperature, °C | 37.9 ± 0.9 | 37.7 ± 0.8 | **0.22** |
| Systolic blood pressure, mmHg | 88.3 ± 15.7 | 88.6 ± 16.6 | 0.02 |
| ECI | 0.07 ± 3.95 | -0.07 ± 1.31 | 0.05 |
| ECI category (>0 %) | ≤11 (≤1.2) | 13 (1.3) | 0.02 |
| Hemoglobin, g/dL | 9.3 ± 2.0 | 8.8 ± 1.8 | **0.27** |
| Missing | 180 (19.8) | 124 (12.7) | **0.19** |
| Total bilirubin, mg/dL | 1.8 ± 5.0 | 1.4 ± 3.5 | 0.10 |
| Missing | 54 (6.0) | 105 (10.8) | **0.17** |
| Creatinine, mg/dL | 1.4 ± 1.3 | 1.9 ± 2.0 | **0.31** |
| Prior steroid use (Yes %) | 722 (79.6) | 663 (67.9) | **0.27** |
| Vasopressor use at $SpO_2$ time (Yes %) | 420 (46.3) | 422 (43.2) | 0.06 |
| Smoking Status (current smoker %) | 26 (2.9) | 36 (3.7) | 0.05 |
| ADI national percentile | 38.1 (23.9) | 53.6 (27.1) | **0.61** |
| ADI national decile | | | **0.76** |
| 1-10 | 126 (13.9) | 53 (5.4) | |
| 11-20 | 113 (12.5) | 86 (8.8) | |
| 21-30 | 176 (19.4) | 98 (10.0) | |
| 31-40 | 125 (13.8) | 57 (5.8) | |
| 41-50 | 91 (10.0) | 187 (19.2) | |
| 51-60 | 41 (4.5) | 57 (5.8) | |
| 61-70 | 127 (14.0) | 146 (15.0) | |
| 71-80 | 59 (6.5) | 61 (6.2) | |
| 81-90 | >30 (>3.3) | 152 (15.6) | |

| | | |
|---|---|---|
| 91-100 | ≤11 (≤1.2) | 79 (8.1) |

Abbreviations: NHW = Non-Hispanic White, NHB = Non-Hispanic Black, No. = Number, BMI = Body Mass Index, ECI = Elixhauser Comorbidity Index, ADI = Area Deprivation Index, SMD = Standardized Mean Difference

Continuous variables are reported as mean ± SD. Categorical variables are presented as frequency (percentage).

For continuous variables, SMDs were calculated as differences in means divided by pooled standard deviations. For categorical variables, SMDs were calculated using a multinomial extension that represents category proportions as a vector and computes a Mahalanobis distance between groups. SMD values > 0.1 are bolded.

Cells with counts ≤11 were suppressed in accordance with CMS policy to protect confidentiality.

**sTable 3. Sensitivity analysis swapping $SaO_2$ and $SpO_2$: simulated additive disparities in dexamethasone treatment and additive intervention effects under varying degrees of racial bias**

| | NHW (N = 907) | NHB (N = 976) | Additive Disparity | Additive intervention effect |
|---|---|---|---|---|
| **% Treated** | | | | |
| Reference (Δμ = 0, Δσ = 0) | 22.2 (16.0, 28.5) | 16.9 (12.2, 21.7) | -5.3 (-11.8, 1.2) | Reference |
| Intervention | | | | |
| Δμ = 2, Δσ = 0 | 22.2 (16.0, 28.5) | 13.4 (8.2, 18.5) | -8.9 (-15.8, -1.9) | -3.5 (-5.7, -1.4) |
| Δμ = 1, Δσ = 0 | 22.2 (16.0, 28.5) | 15.4 (10.5, 20.2) | -6.9 (-13.5, -0.2) | -1.5 (-2.6, -0.5) |
| Δμ = 0, Δσ = 3 | 22.2 (16.0, 28.5) | 12.9 (8.2, 17.7) | -9.3 (-15.9, -2.7) | -4.0 (-6.4, -1.6) |
| Δμ = 0, Δσ = 1.5 | 22.2 (16.0, 28.5) | 14.9 (10.3, 19.5) | -7.3 (-13.8, -0.9) | -2.0 (-3.3, -0.7) |
| Δμ = 2, Δσ = 3 | 22.2 (16.0, 28.5) | 11.6 (6.8, 16.3) | -10.7 (-17.4, -4.0) | -5.4 (-7.9, -2.8) |
| Δμ = 2, Δσ = 1.5 | 22.2 (16.0, 28.5) | 12.6 (7.8, 17.5) | -9.6 (-16.3, -2.9) | -4.3 (-6.4, -2.2) |
| Δμ = 1, Δσ = 3 | 22.2 (16.0, 28.5) | 12.4 (7.6, 17.1) | -9.9 (-16.5, -3.3) | -4.6 (-6.9, -2.2) |
| Δμ = 1, Δσ = 1.5 | 22.2 (16.0, 28.5) | 13.9 (9.3, 18.6) | -8.3 (-14.8, -1.8) | -3.0 (-4.4, -1.5) |
| **RMST, minutes** | | | | |
| Reference (Δμ = 0, Δσ = 0) | 1251.5 (1190.4, 1312.6) | 1290.3 (1215.9, 1364.8) | 38.9 (-51.9, 129.6) | Reference |
| Intervention | | | | |
| Δμ = 2, Δσ = 0 | 1251.5 (1190.4, 1312.6) | 1302.9 (1174.1, 1431.8) | 51.5 (-89.3, 192.2) | 12.6 (-48.7, 73.9) |
| Δμ = 1, Δσ = 0 | 1251.5 (1190.4, 1312.6) | 1295.8 (1197.3, 1394.3) | 44.3 (-68.0, 156.7) | 5.5 (-22.6, 33.5) |
| Δμ = 0, Δσ = 3 | 1251.5 (1190.4, 1312.6) | 1290.2 (1219.4, 1360.9) | 38.7 (-48.9, 126.2) | -0.2 (-9.7, 9.3) |
| Δμ = 0, Δσ = 1.5 | 1251.5 (1190.4, 1312.6) | 1288.3 (1217.1, 1359.6) | 36.9 (-51.0, 124.8) | -2.0 (-8.7, 4.7) |
| Δμ = 2, Δσ = 3 | 1251.5 (1190.4, 1312.6) | 1290.8 (1211.1, 1370.5) | 39.3 (-56.0, 134.6) | 0.5 (-7.9, 8.8) |
| Δμ = 2, Δσ = 1.5 | 1251.5 (1190.4, 1312.6) | 1292.1 (1202.2, 1381.9) | 40.6 (-63.8, 144.9) | 1.7 (-16.1, 19.5) |
| Δμ = 1, Δσ = 3 | 1251.5 (1190.4, 1312.6) | 1290.3 (1215.2, 1365.4) | 38.8 (-52.4, 130.1) | 0.0 (-7.1, 7.1) |
| Δμ = 1, Δσ = 1.5 | 1251.5 (1190.4, 1312.6) | 1289.8 (1209.9, 1369.8) | 38.4 (-57.1, 133.8) | -0.5 (-7.6, 6.6) |

Abbreviations: NHW = Non-Hispanic White, NHB = Non-Hispanic Black, RMST = Restricted Mean Survival Time

Note: Values are presented as point estimates, with 95% confidence intervals shown in parentheses.

**sTable 4. Sensitivity analysis excluding prior-window $SaO_2$ measurements: simulated additive disparities in dexamethasone treatment and additive intervention effects under varying degrees of racial bias**

| | NHW (N = 730) | NHB (N = 824) | Additive disparity | Additive intervention effect |
|---|---|---|---|---|
| **% Treated** | | | | |
| Reference (Δμ = 0, Δσ = 0) | 22.7 (17.1, 28.3) | 19.2 (14.6, 23.8) | -3.5 (-9.6, 2.6) | Reference |
| Intervention | | | | |
| Δμ = 2, Δσ = 0 | 22.7 (17.1, 28.3) | 20.5 (16.1, 24.9) | -2.2 (-7.9, 3.6) | 1.3 (-0.7, 3.4) |
| Δμ = 1, Δσ = 0 | 22.7 (17.1, 28.3) | 19.8 (15.4, 24.2) | -2.8 (-8.7, 3.0) | 0.6 (-0.5, 1.8) |
| Δμ = 0, Δσ = 3 | 22.7 (17.1, 28.3) | 19.5 (14.0, 25.0) | -3.2 (-10.0, 3.6) | 0.3 (-2.6, 3.3) |
| Δμ = 0, Δσ = 1.5 | 22.7 (17.1, 28.3) | 19.4 (14.6, 24.1) | -3.3 (-9.6, 2.9) | 0.2 (-1.1, 1.5) |
| Δμ = 2, Δσ = 3 | 22.7 (17.1, 28.3) | 20.7 (15.9, 25.5) | -2.0 (-8.0, 4.1) | 1.5 (-1.0, 4.0) |
| Δμ = 2, Δσ = 1.5 | 22.7 (17.1, 28.3) | 20.6 (16.3, 25.0) | -2.0 (-7.7, 3.7) | 1.5 (-0.3, 3.3) |
| Δμ = 1, Δσ = 3 | 22.7 (17.1, 28.3) | 20.1 (15.1, 25.1) | -2.6 (-8.9, 3.8) | 0.9 (-1.5, 3.4) |
| Δμ = 1, Δσ = 1.5 | 22.7 (17.1, 28.3) | 20.0 (15.5, 24.4) | -2.7 (-8.6, 3.2) | 0.8 (-0.4, 2.0) |
| **RMST, minutes** | | | | |
| Reference (Δμ = 0, Δσ = 0) | 1262.9 (1208.9, 1316.8) | 1304.9 (1258.3, 1351.6) | 42.1 (-24.3, 108.4) | Reference |
| Intervention | | | | |
| Δμ = 2, Δσ = 0 | 1262.9 (1208.9, 1316.8) | 1302.2 (1258.4, 1345.9) | 39.3 (-24.7, 103.3) | -2.8 (-8.6, 3.0) |
| Δμ = 1, Δσ = 0 | 1262.9 (1208.9, 1316.8) | 1303.4 (1258.2, 1348.6) | 40.5 (-24.7, 105.8) | -1.5 (-4.4, 1.3) |
| Δμ = 0, Δσ = 3 | 1262.9 (1208.9, 1316.8) | 1298.3 (1257.1, 1339.6) | 35.4 (-26.3, 97.1) | -6.6 (-23.7, 10.4) |
| Δμ = 0, Δσ = 1.5 | 1262.9 (1208.9, 1316.8) | 1299.5 (1257.8, 1341.3) | 36.6 (-25.7, 99.0) | -5.4 (-16.2, 5.4) |
| Δμ = 2, Δσ = 3 | 1262.9 (1208.9, 1316.8) | 1297.5 (1255.4, 1339.5) | 34.6 (-27.4, 96.5) | -7.5 (30.5, 15.5) |
| Δμ = 2, Δσ = 1.5 | 1262.9 (1208.9, 1316.8) | 1298.6 (1257.3, 1339.9) | 35.7 (-26.1, 97.5) | -6.4 (-22.3, 9.5) |
| Δμ = 1, Δσ = 3 | 1262.9 (1208.9, 1316.8) | 1297.9 (1256.4, 1339.5) | 35.1 (-26.7, 96.8) | -7.0 (-26.8, 12.8) |
| Δμ = 1, Δσ = 1.5 | 1262.9 (1208.9, 1316.8) | 1299.0 (1257.7, 1340.4) | 36.2 (-25.8, 98.1) | -5.9 (-18.8, 7.0) |

Abbreviations: NHW = Non-Hispanic White, NHB = Non-Hispanic Black, RMST = Restricted Mean Survival Time

Note: Values are presented as point estimates, with 95% confidence intervals shown in parentheses.

**sTable 5. Natural course additive disparities in 24-hour dexamethasone treatment receipt (%) and time-to-treatment**

| | NHW (N = 907) | NHB (N = 976) | Additive disparity |
|---|---|---|---|
| % Treated | | | |
| Observed | 26.7 (21.3, 32.1) | 20.3 (15.8, 24.8) | -6.4 (-13.4, 0.6) |
| Natural course | 26.7 (21.4, 32.0) | 20.3 (15.7, 24.8) | -6.4 (-13.5, 0.7) |
| Observed (standardized) | 22.5 (16.7, 28.3) | 20.3 (15.8, 24.8) | -2.2 (-8.2, 3.7) |
| Natural course (standardized) | 22.5 (16.6, 28.3) | 20.3 (15.7, 24.8) | -2.2 (-7.8, 3.5) |
| RMST, minutes | | | |
| Observed | 1231.2 (1188.2, 1274.3) | 1285.5 (1245.7, 1325.3) | 54.2 (-5.5, 114.0) |
| Natural course | 1236.2 (1189.7, 1282.7) | 1295.6 (1259.1, 1332.1) | 59.4 (-0.1, 118.9) |
| Observed (standardized) | 1269.9 (1220.4, 1319.4) | 1293.5 (1254.9, 1332.2) | 23.6 (-27.0, 74.2) |
| Natural course (standardized) | 1268.8 (1220.2, 1317.4) | 1295.6 (1259.1, 1332.1) | 26.9 (-20.9, 74.7) |

Abbreviations: NHW = Non-Hispanic White, NHB = Non-Hispanic Black, RMST = Restricted Mean Survival Time

Note: Values are presented as point estimates, with 95% confidence intervals shown in parentheses.

**sTable 6. Summary of missingness of covariates**

| Variable | Missing count | Missing percentage |
|---|---|---|
| $SaO_2$ | 0 | 0 |
| Age | 0 | 0 |
| Sex | 0 | 0 |
| ECI | 0 | 0 |
| Prior steroid use | 0 | 0 |
| Vasopressor use at $SpO_2$ time | 0 | 0 |
| BMI | 2186 | 20.9 |
| Pulse rate | 143 | 1.4 |
| Respiratory rate | 173 | 1.7 |
| Body temperature | 169 | 1.6 |
| Systolic blood pressure | 224 | 2.1 |
| Hemoglobin, g/dL | 3111 | 29.8 |
| Total bilirubin, mg/dL | 899 | 8.6 |
| Creatinine, mg/dL | 5 | 0.05 |
| Smoking status | 917 | 8.8 |
| ADI | 2909 | 27.8 |

Abbreviations: ECI = Elixhauser Comorbidity Index, BMI = Body Mass Index; ADI = Area Deprivation Index

**sTable 7. Number of participants in each trial indexed by calendar time**

| Trial number | Year | Half-day | Calendar time | No. Participants |
|---|---|---|---|---|
| 1 | 2020 | 165 | 2020-03-23 12:00:00 UTC | 3 |
| 2 | 2020 | 172 | 2020-03-27 00:00:00 UTC | 2 |
| 3 | 2020 | 174 | 2020-03-28 00:00:00 UTC | 3 |
| 4 | 2020 | 179 | 2020-03-30 12:00:00 UTC | 4 |
| 5 | 2020 | 181 | 2020-03-31 12:00:00 UTC | 6 |
| 6 | 2020 | 183 | 2020-04-01 12:00:00 UTC | 6 |
| 7 | 2020 | 185 | 2020-04-02 12:00:00 UTC | 5 |
| 8 | 2020 | 186 | 2020-04-03 00:00:00 UTC | 2 |
| 9 | 2020 | 189 | 2020-04-04 12:00:00 UTC | 2 |
| 10 | 2020 | 191 | 2020-04-05 12:00:00 UTC | 4 |
| 11 | 2020 | 192 | 2020-04-06 00:00:00 UTC | 3 |
| 12 | 2020 | 194 | 2020-04-07 00:00:00 UTC | 7 |
| 13 | 2020 | 195 | 2020-04-07 12:00:00 UTC | 6 |
| 14 | 2020 | 197 | 2020-04-08 12:00:00 UTC | 8 |
| 15 | 2020 | 199 | 2020-04-09 12:00:00 UTC | 4 |
| 16 | 2020 | 200 | 2020-04-10 00:00:00 UTC | 4 |
| 17 | 2020 | 201 | 2020-04-10 12:00:00 UTC | 3 |
| 18 | 2020 | 202 | 2020-04-11 00:00:00 UTC | 5 |
| 19 | 2020 | 203 | 2020-04-11 12:00:00 UTC | 5 |
| 20 | 2020 | 204 | 2020-04-12 00:00:00 UTC | 3 |
| 21 | 2020 | 205 | 2020-04-12 12:00:00 UTC | 6 |
| 22 | 2020 | 206 | 2020-04-13 00:00:00 UTC | 3 |
| 23 | 2020 | 209 | 2020-04-14 12:00:00 UTC | 5 |
| 24 | 2020 | 210 | 2020-04-15 00:00:00 UTC | 8 |
| 25 | 2020 | 211 | 2020-04-15 12:00:00 UTC | 6 |
| 26 | 2020 | 212 | 2020-04-16 00:00:00 UTC | 7 |
| 27 | 2020 | 213 | 2020-04-16 12:00:00 UTC | 6 |
| 28 | 2020 | 214 | 2020-04-17 00:00:00 UTC | 8 |
| 29 | 2020 | 215 | 2020-04-17 12:00:00 UTC | 4 |
| 30 | 2020 | 216 | 2020-04-18 00:00:00 UTC | 5 |
| 31 | 2020 | 217 | 2020-04-18 12:00:00 UTC | 5 |
| 32 | 2020 | 218 | 2020-04-19 00:00:00 UTC | 5 |
| 33 | 2020 | 219 | 2020-04-19 12:00:00 UTC | 7 |
| 34 | 2020 | 222 | 2020-04-21 00:00:00 UTC | 3 |
| 35 | 2020 | 223 | 2020-04-21 12:00:00 UTC | 5 |
| 36 | 2020 | 224 | 2020-04-22 00:00:00 UTC | 4 |
| 37 | 2020 | 225 | 2020-04-22 12:00:00 UTC | 5 |
| 38 | 2020 | 226 | 2020-04-23 00:00:00 UTC | 4 |
| 39 | 2020 | 228 | 2020-04-24 00:00:00 UTC | 10 |
| 40 | 2020 | 229 | 2020-04-24 12:00:00 UTC | 7 |
| 41 | 2020 | 230 | 2020-04-25 00:00:00 UTC | 5 |
| 42 | 2020 | 231 | 2020-04-25 12:00:00 UTC | 7 |
| 43 | 2020 | 232 | 2020-04-26 00:00:00 UTC | 10 |
| 44 | 2020 | 233 | 2020-04-26 12:00:00 UTC | 7 |
| 45 | 2020 | 234 | 2020-04-27 00:00:00 UTC | 5 |
| 46 | 2020 | 235 | 2020-04-27 12:00:00 UTC | 6 |
| 47 | 2020 | 236 | 2020-04-28 00:00:00 UTC | 5 |
| 48 | 2020 | 238 | 2020-04-29 00:00:00 UTC | 4 |
| 49 | 2020 | 239 | 2020-04-29 12:00:00 UTC | 7 |
| 50 | 2020 | 240 | 2020-04-30 00:00:00 UTC | 6 |
| 51 | 2020 | 241 | 2020-04-30 12:00:00 UTC | 3 |
| 52 | 2020 | 242 | 2020-05-01 00:00:00 UTC | 3 |
| 53 | 2020 | 243 | 2020-05-01 12:00:00 UTC | 6 |
| 54 | 2020 | 244 | 2020-05-02 00:00:00 UTC | 7 |
| 55 | 2020 | 245 | 2020-05-02 12:00:00 UTC | 9 |
| 56 | 2020 | 246 | 2020-05-03 00:00:00 UTC | 6 |
| 57 | 2020 | 247 | 2020-05-03 12:00:00 UTC | 5 |
| 58 | 2020 | 248 | 2020-05-04 00:00:00 UTC | 3 |
| 59 | 2020 | 250 | 2020-05-05 00:00:00 UTC | 5 |
| 60 | 2020 | 252 | 2020-05-06 00:00:00 UTC | 2 |
| 61 | 2020 | 253 | 2020-05-06 12:00:00 UTC | 5 |
| 62 | 2020 | 254 | 2020-05-07 00:00:00 UTC | 4 |
| 63 | 2020 | 255 | 2020-05-07 12:00:00 UTC | 6 |
| 64 | 2020 | 256 | 2020-05-08 00:00:00 UTC | 5 |
| 65 | 2020 | 258 | 2020-05-09 00:00:00 UTC | 7 |
| 66 | 2020 | 259 | 2020-05-09 12:00:00 UTC | 11 |
| 67 | 2020 | 260 | 2020-05-10 00:00:00 UTC | 4 |
| 68 | 2020 | 261 | 2020-05-10 12:00:00 UTC | 5 |
| 69 | 2020 | 262 | 2020-05-11 00:00:00 UTC | 7 |
| 70 | 2020 | 263 | 2020-05-11 12:00:00 UTC | 6 |
| 71 | 2020 | 264 | 2020-05-12 00:00:00 UTC | 2 |
| 72 | 2020 | 265 | 2020-05-12 12:00:00 UTC | 4 |
| 73 | 2020 | 266 | 2020-05-13 00:00:00 UTC | 4 |
| 74 | 2020 | 267 | 2020-05-13 12:00:00 UTC | 6 |
| 75 | 2020 | 269 | 2020-05-14 12:00:00 UTC | 5 |
| 76 | 2020 | 270 | 2020-05-15 00:00:00 UTC | 6 |
| 77 | 2020 | 271 | 2020-05-15 12:00:00 UTC | 8 |
| 78 | 2020 | 272 | 2020-05-16 00:00:00 UTC | 6 |
| 79 | 2020 | 274 | 2020-05-17 00:00:00 UTC | 4 |
| 80 | 2020 | 275 | 2020-05-17 12:00:00 UTC | 7 |
| 81 | 2020 | 280 | 2020-05-20 00:00:00 UTC | 7 |
| 82 | 2020 | 281 | 2020-05-20 12:00:00 UTC | 5 |
| 83 | 2020 | 282 | 2020-05-21 00:00:00 UTC | 4 |
| 84 | 2020 | 283 | 2020-05-21 12:00:00 UTC | 3 |
| 85 | 2020 | 284 | 2020-05-22 00:00:00 UTC | 5 |
| 86 | 2020 | 287 | 2020-05-23 12:00:00 UTC | 2 |
| 87 | 2020 | 288 | 2020-05-24 00:00:00 UTC | 3 |
| 88 | 2020 | 292 | 2020-05-26 00:00:00 UTC | 3 |
| 89 | 2020 | 295 | 2020-05-27 12:00:00 UTC | 4 |
| 90 | 2020 | 296 | 2020-05-28 00:00:00 UTC | 4 |
| 91 | 2020 | 297 | 2020-05-28 12:00:00 UTC | 4 |
| 92 | 2020 | 299 | 2020-05-29 12:00:00 UTC | 4 |
| 93 | 2020 | 300 | 2020-05-30 00:00:00 UTC | 3 |
| 94 | 2020 | 309 | 2020-06-03 12:00:00 UTC | 8 |
| 95 | 2020 | 315 | 2020-06-06 12:00:00 UTC | 5 |
| 96 | 2020 | 316 | 2020-06-07 00:00:00 UTC | 2 |
| 97 | 2020 | 321 | 2020-06-09 12:00:00 UTC | 2 |
| 98 | 2020 | 325 | 2020-06-11 12:00:00 UTC | 2 |
| 99 | 2020 | 340 | 2020-06-19 00:00:00 UTC | 4 |
| 100 | 2020 | 347 | 2020-06-22 12:00:00 UTC | 5 |
| 101 | 2020 | 390 | 2020-07-14 00:00:00 UTC | 2 |
| 102 | 2020 | 392 | 2020-07-15 00:00:00 UTC | 2 |
| 103 | 2020 | 405 | 2020-07-21 12:00:00 UTC | 2 |
| 104 | 2020 | 409 | 2020-07-23 12:00:00 UTC | 2 |
| 105 | 2020 | 552 | 2020-10-03 00:00:00 UTC | 2 |
| 106 | 2020 | 560 | 2020-10-07 00:00:00 UTC | 2 |
| 107 | 2020 | 566 | 2020-10-10 00:00:00 UTC | 2 |
| 108 | 2020 | 632 | 2020-11-12 00:00:00 UTC | 2 |
| 109 | 2020 | 642 | 2020-11-17 00:00:00 UTC | 2 |
| 110 | 2020 | 643 | 2020-11-17 12:00:00 UTC | 2 |
| 111 | 2020 | 647 | 2020-11-19 12:00:00 UTC | 2 |
| 112 | 2020 | 648 | 2020-11-20 00:00:00 UTC | 3 |
| 113 | 2020 | 652 | 2020-11-22 00:00:00 UTC | 2 |
| 114 | 2020 | 655 | 2020-11-23 12:00:00 UTC | 2 |
| 115 | 2020 | 657 | 2020-11-24 12:00:00 UTC | 3 |
| 116 | 2020 | 658 | 2020-11-25 00:00:00 UTC | 4 |
| 117 | 2020 | 660 | 2020-11-26 00:00:00 UTC | 4 |
| 118 | 2020 | 663 | 2020-11-27 12:00:00 UTC | 2 |
| 119 | 2020 | 665 | 2020-11-28 12:00:00 UTC | 2 |
| 120 | 2020 | 667 | 2020-11-29 12:00:00 UTC | 4 |
| 121 | 2020 | 669 | 2020-11-30 12:00:00 UTC | 4 |
| 122 | 2020 | 671 | 2020-12-01 12:00:00 UTC | 4 |
| 123 | 2020 | 673 | 2020-12-02 12:00:00 UTC | 4 |
| 124 | 2020 | 675 | 2020-12-03 12:00:00 UTC | 5 |
| 125 | 2020 | 676 | 2020-12-04 00:00:00 UTC | 2 |
| 126 | 2020 | 678 | 2020-12-05 00:00:00 UTC | 4 |
| 127 | 2020 | 679 | 2020-12-05 12:00:00 UTC | 3 |
| 128 | 2020 | 680 | 2020-12-06 00:00:00 UTC | 5 |
| 129 | 2020 | 684 | 2020-12-08 00:00:00 UTC | 3 |
| 130 | 2020 | 686 | 2020-12-09 00:00:00 UTC | 2 |
| 131 | 2020 | 687 | 2020-12-09 12:00:00 UTC | 5 |
| 132 | 2020 | 688 | 2020-12-10 00:00:00 UTC | 4 |
| 133 | 2020 | 689 | 2020-12-10 12:00:00 UTC | 3 |
| 134 | 2020 | 691 | 2020-12-11 12:00:00 UTC | 3 |
| 135 | 2020 | 694 | 2020-12-13 00:00:00 UTC | 4 |
| 136 | 2020 | 695 | 2020-12-13 12:00:00 UTC | 6 |
| 137 | 2020 | 697 | 2020-12-14 12:00:00 UTC | 11 |
| 138 | 2020 | 698 | 2020-12-15 00:00:00 UTC | 4 |
| 139 | 2020 | 699 | 2020-12-15 12:00:00 UTC | 3 |
| 140 | 2020 | 700 | 2020-12-16 00:00:00 UTC | 7 |
| 141 | 2020 | 701 | 2020-12-16 12:00:00 UTC | 3 |
| 142 | 2020 | 702 | 2020-12-17 00:00:00 UTC | 5 |
| 143 | 2020 | 704 | 2020-12-18 00:00:00 UTC | 6 |
| 144 | 2020 | 705 | 2020-12-18 12:00:00 UTC | 7 |
| 145 | 2020 | 706 | 2020-12-19 00:00:00 UTC | 5 |
| 146 | 2020 | 707 | 2020-12-19 12:00:00 UTC | 5 |
| 147 | 2020 | 708 | 2020-12-20 00:00:00 UTC | 8 |
| 148 | 2020 | 709 | 2020-12-20 12:00:00 UTC | 7 |
| 149 | 2020 | 710 | 2020-12-21 00:00:00 UTC | 6 |
| 150 | 2020 | 711 | 2020-12-21 12:00:00 UTC | 6 |
| 151 | 2020 | 712 | 2020-12-22 00:00:00 UTC | 5 |
| 152 | 2020 | 713 | 2020-12-22 12:00:00 UTC | 4 |
| 153 | 2020 | 714 | 2020-12-23 00:00:00 UTC | 5 |
| 154 | 2020 | 715 | 2020-12-23 12:00:00 UTC | 4 |
| 155 | 2020 | 716 | 2020-12-24 00:00:00 UTC | 7 |
| 156 | 2020 | 717 | 2020-12-24 12:00:00 UTC | 6 |
| 157 | 2020 | 718 | 2020-12-25 00:00:00 UTC | 5 |
| 158 | 2020 | 719 | 2020-12-25 12:00:00 UTC | 4 |
| 159 | 2020 | 721 | 2020-12-26 12:00:00 UTC | 8 |
| 160 | 2020 | 722 | 2020-12-27 00:00:00 UTC | 5 |
| 161 | 2020 | 724 | 2020-12-28 00:00:00 UTC | 10 |
| 162 | 2020 | 725 | 2020-12-28 12:00:00 UTC | 4 |
| 163 | 2020 | 726 | 2020-12-29 00:00:00 UTC | 7 |
| 164 | 2020 | 727 | 2020-12-29 12:00:00 UTC | 5 |
| 165 | 2020 | 728 | 2020-12-30 00:00:00 UTC | 10 |
| 166 | 2020 | 729 | 2020-12-30 12:00:00 UTC | 6 |
| 167 | 2020 | 730 | 2020-12-31 00:00:00 UTC | 4 |
| 168 | 2021 | 1 | 2021-01-01 12:00:00 UTC | 6 |
| 169 | 2021 | 2 | 2021-01-02 00:00:00 UTC | 8 |
| 170 | 2021 | 3 | 2021-01-02 12:00:00 UTC | 6 |
| 171 | 2021 | 4 | 2021-01-03 00:00:00 UTC | 6 |
| 172 | 2021 | 5 | 2021-01-03 12:00:00 UTC | 6 |
| 173 | 2021 | 6 | 2021-01-04 00:00:00 UTC | 4 |
| 174 | 2021 | 7 | 2021-01-04 12:00:00 UTC | 6 |
| 175 | 2021 | 8 | 2021-01-05 00:00:00 UTC | 2 |
| 176 | 2021 | 9 | 2021-01-05 12:00:00 UTC | 3 |
| 177 | 2021 | 11 | 2021-01-06 12:00:00 UTC | 4 |
| 178 | 2021 | 14 | 2021-01-08 00:00:00 UTC | 3 |
| 179 | 2021 | 15 | 2021-01-08 12:00:00 UTC | 6 |
| 180 | 2021 | 16 | 2021-01-09 00:00:00 UTC | 7 |
| 181 | 2021 | 17 | 2021-01-09 12:00:00 UTC | 4 |
| 182 | 2021 | 18 | 2021-01-10 00:00:00 UTC | 7 |
| 183 | 2021 | 19 | 2021-01-10 12:00:00 UTC | 5 |
| 184 | 2021 | 20 | 2021-01-11 00:00:00 UTC | 4 |
| 185 | 2021 | 21 | 2021-01-11 12:00:00 UTC | 6 |
| 186 | 2021 | 23 | 2021-01-12 12:00:00 UTC | 7 |
| 187 | 2021 | 24 | 2021-01-13 00:00:00 UTC | 7 |
| 188 | 2021 | 25 | 2021-01-13 12:00:00 UTC | 5 |
| 189 | 2021 | 26 | 2021-01-14 00:00:00 UTC | 7 |
| 190 | 2021 | 27 | 2021-01-14 12:00:00 UTC | 5 |
| 191 | 2021 | 28 | 2021-01-15 00:00:00 UTC | 7 |
| 192 | 2021 | 29 | 2021-01-15 12:00:00 UTC | 4 |
| 193 | 2021 | 30 | 2021-01-16 00:00:00 UTC | 4 |
| 194 | 2021 | 31 | 2021-01-16 12:00:00 UTC | 9 |
| 195 | 2021 | 32 | 2021-01-17 00:00:00 UTC | 10 |
| 196 | 2021 | 33 | 2021-01-17 12:00:00 UTC | 5 |
| 197 | 2021 | 35 | 2021-01-18 12:00:00 UTC | 8 |
| 198 | 2021 | 36 | 2021-01-19 00:00:00 UTC | 4 |
| 199 | 2021 | 37 | 2021-01-19 12:00:00 UTC | 5 |
| 200 | 2021 | 38 | 2021-01-20 00:00:00 UTC | 4 |
| 201 | 2021 | 39 | 2021-01-20 12:00:00 UTC | 10 |
| 202 | 2021 | 40 | 2021-01-21 00:00:00 UTC | 10 |
| 203 | 2021 | 41 | 2021-01-21 12:00:00 UTC | 4 |
| 204 | 2021 | 43 | 2021-01-22 12:00:00 UTC | 5 |
| 205 | 2021 | 45 | 2021-01-23 12:00:00 UTC | 6 |
| 206 | 2021 | 46 | 2021-01-24 00:00:00 UTC | 6 |
| 207 | 2021 | 47 | 2021-01-24 12:00:00 UTC | 3 |
| 208 | 2021 | 48 | 2021-01-25 00:00:00 UTC | 4 |
| 209 | 2021 | 49 | 2021-01-25 12:00:00 UTC | 4 |
| 210 | 2021 | 50 | 2021-01-26 00:00:00 UTC | 3 |
| 211 | 2021 | 51 | 2021-01-26 12:00:00 UTC | 6 |
| 212 | 2021 | 52 | 2021-01-27 00:00:00 UTC | 5 |
| 213 | 2021 | 53 | 2021-01-27 12:00:00 UTC | 10 |
| 214 | 2021 | 54 | 2021-01-28 00:00:00 UTC | 3 |
| 215 | 2021 | 55 | 2021-01-28 12:00:00 UTC | 5 |
| 216 | 2021 | 56 | 2021-01-29 00:00:00 UTC | 6 |
| 217 | 2021 | 57 | 2021-01-29 12:00:00 UTC | 5 |
| 218 | 2021 | 58 | 2021-01-30 00:00:00 UTC | 3 |
| 219 | 2021 | 60 | 2021-01-31 00:00:00 UTC | 4 |
| 220 | 2021 | 61 | 2021-01-31 12:00:00 UTC | 5 |
| 221 | 2021 | 63 | 2021-02-01 12:00:00 UTC | 3 |
| 222 | 2021 | 64 | 2021-02-02 00:00:00 UTC | 3 |
| 223 | 2021 | 66 | 2021-02-03 00:00:00 UTC | 2 |
| 224 | 2021 | 67 | 2021-02-03 12:00:00 UTC | 4 |
| 225 | 2021 | 68 | 2021-02-04 00:00:00 UTC | 3 |
| 226 | 2021 | 69 | 2021-02-04 12:00:00 UTC | 3 |
| 227 | 2021 | 73 | 2021-02-06 12:00:00 UTC | 7 |
| 228 | 2021 | 75 | 2021-02-07 12:00:00 UTC | 3 |
| 229 | 2021 | 77 | 2021-02-08 12:00:00 UTC | 4 |
| 230 | 2021 | 78 | 2021-02-09 00:00:00 UTC | 4 |
| 231 | 2021 | 79 | 2021-02-09 12:00:00 UTC | 8 |
| 232 | 2021 | 80 | 2021-02-10 00:00:00 UTC | 3 |
| 233 | 2021 | 81 | 2021-02-10 12:00:00 UTC | 4 |
| 234 | 2021 | 82 | 2021-02-11 00:00:00 UTC | 3 |
| 235 | 2021 | 83 | 2021-02-11 12:00:00 UTC | 3 |
| 236 | 2021 | 84 | 2021-02-12 00:00:00 UTC | 4 |
| 237 | 2021 | 85 | 2021-02-12 12:00:00 UTC | 3 |
| 238 | 2021 | 87 | 2021-02-13 12:00:00 UTC | 4 |
| 239 | 2021 | 88 | 2021-02-14 00:00:00 UTC | 9 |
| 240 | 2021 | 90 | 2021-02-15 00:00:00 UTC | 4 |
| 241 | 2021 | 92 | 2021-02-16 00:00:00 UTC | 3 |
| 242 | 2021 | 93 | 2021-02-16 12:00:00 UTC | 4 |
| 243 | 2021 | 94 | 2021-02-17 00:00:00 UTC | 2 |
| 244 | 2021 | 95 | 2021-02-17 12:00:00 UTC | 2 |
| 245 | 2021 | 96 | 2021-02-18 00:00:00 UTC | 3 |
| 246 | 2021 | 97 | 2021-02-18 12:00:00 UTC | 3 |
| 247 | 2021 | 98 | 2021-02-19 00:00:00 UTC | 4 |
| 248 | 2021 | 99 | 2021-02-19 12:00:00 UTC | 7 |
| 249 | 2021 | 100 | 2021-02-20 00:00:00 UTC | 5 |
| 250 | 2021 | 101 | 2021-02-20 12:00:00 UTC | 4 |
| 251 | 2021 | 102 | 2021-02-21 00:00:00 UTC | 3 |
| 252 | 2021 | 107 | 2021-02-23 12:00:00 UTC | 3 |
| 253 | 2021 | 110 | 2021-02-25 00:00:00 UTC | 4 |
| 254 | 2021 | 116 | 2021-02-28 00:00:00 UTC | 3 |
| 255 | 2021 | 118 | 2021-03-01 00:00:00 UTC | 3 |
| 256 | 2021 | 122 | 2021-03-03 00:00:00 UTC | 3 |
| 257 | 2021 | 126 | 2021-03-05 00:00:00 UTC | 2 |
| 258 | 2021 | 128 | 2021-03-06 00:00:00 UTC | 3 |
| 259 | 2021 | 133 | 2021-03-08 12:00:00 UTC | 2 |
| 260 | 2021 | 141 | 2021-03-12 12:00:00 UTC | 3 |
| 261 | 2021 | 142 | 2021-03-13 00:00:00 UTC | 2 |
| 262 | 2021 | 143 | 2021-03-13 12:00:00 UTC | 3 |
| 263 | 2021 | 146 | 2021-03-15 00:00:00 UTC | 5 |
| 264 | 2021 | 147 | 2021-03-15 12:00:00 UTC | 5 |
| 265 | 2021 | 149 | 2021-03-16 12:00:00 UTC | 3 |
| 266 | 2021 | 160 | 2021-03-22 00:00:00 UTC | 3 |
| 267 | 2021 | 163 | 2021-03-23 12:00:00 UTC | 2 |
| 268 | 2021 | 166 | 2021-03-25 00:00:00 UTC | 2 |
| 269 | 2021 | 167 | 2021-03-25 12:00:00 UTC | 2 |
| 270 | 2021 | 169 | 2021-03-26 12:00:00 UTC | 2 |
| 271 | 2021 | 170 | 2021-03-27 00:00:00 UTC | 4 |
| 272 | 2021 | 173 | 2021-03-28 12:00:00 UTC | 3 |
| 273 | 2021 | 174 | 2021-03-29 00:00:00 UTC | 4 |
| 274 | 2021 | 176 | 2021-03-30 00:00:00 UTC | 3 |
| 275 | 2021 | 177 | 2021-03-30 12:00:00 UTC | 3 |
| 276 | 2021 | 179 | 2021-03-31 12:00:00 UTC | 2 |
| 277 | 2021 | 182 | 2021-04-02 00:00:00 UTC | 4 |
| 278 | 2021 | 183 | 2021-04-02 12:00:00 UTC | 3 |
| 279 | 2021 | 185 | 2021-04-03 12:00:00 UTC | 2 |
| 280 | 2021 | 188 | 2021-04-05 00:00:00 UTC | 6 |
| 281 | 2021 | 190 | 2021-04-06 00:00:00 UTC | 4 |
| 282 | 2021 | 191 | 2021-04-06 12:00:00 UTC | 3 |
| 283 | 2021 | 193 | 2021-04-07 12:00:00 UTC | 2 |
| 284 | 2021 | 194 | 2021-04-08 00:00:00 UTC | 3 |
| 285 | 2021 | 197 | 2021-04-09 12:00:00 UTC | 2 |
| 286 | 2021 | 198 | 2021-04-10 00:00:00 UTC | 2 |
| 287 | 2021 | 207 | 2021-04-14 12:00:00 UTC | 2 |
| 288 | 2021 | 208 | 2021-04-15 00:00:00 UTC | 2 |
| 289 | 2021 | 211 | 2021-04-16 12:00:00 UTC | 2 |
| 290 | 2021 | 213 | 2021-04-17 12:00:00 UTC | 2 |
| 291 | 2021 | 215 | 2021-04-18 12:00:00 UTC | 2 |
| 292 | 2021 | 217 | 2021-04-19 12:00:00 UTC | 3 |
| 293 | 2021 | 219 | 2021-04-20 12:00:00 UTC | 5 |
| 294 | 2021 | 220 | 2021-04-21 00:00:00 UTC | 2 |
| 295 | 2021 | 221 | 2021-04-21 12:00:00 UTC | 4 |
| 296 | 2021 | 222 | 2021-04-22 00:00:00 UTC | 3 |
| 297 | 2021 | 227 | 2021-04-24 12:00:00 UTC | 2 |
| 298 | 2021 | 228 | 2021-04-25 00:00:00 UTC | 2 |
| 299 | 2021 | 229 | 2021-04-25 12:00:00 UTC | 5 |
| 300 | 2021 | 232 | 2021-04-27 00:00:00 UTC | 6 |
| 301 | 2021 | 236 | 2021-04-29 00:00:00 UTC | 3 |
| 302 | 2021 | 238 | 2021-04-30 00:00:00 UTC | 4 |
| 303 | 2021 | 240 | 2021-05-01 00:00:00 UTC | 4 |
| 304 | 2021 | 242 | 2021-05-02 00:00:00 UTC | 2 |
| 305 | 2021 | 245 | 2021-05-03 12:00:00 UTC | 4 |
| 306 | 2021 | 246 | 2021-05-04 00:00:00 UTC | 2 |
| 307 | 2021 | 247 | 2021-05-04 12:00:00 UTC | 2 |
| 308 | 2021 | 248 | 2021-05-05 00:00:00 UTC | 2 |
| 309 | 2021 | 249 | 2021-05-05 12:00:00 UTC | 4 |
| 310 | 2021 | 252 | 2021-05-07 00:00:00 UTC | 3 |
| 311 | 2021 | 253 | 2021-05-07 12:00:00 UTC | 3 |
| 312 | 2021 | 258 | 2021-05-10 00:00:00 UTC | 2 |
| 313 | 2021 | 262 | 2021-05-12 00:00:00 UTC | 3 |
| 314 | 2021 | 264 | 2021-05-13 00:00:00 UTC | 4 |
| 315 | 2021 | 266 | 2021-05-14 00:00:00 UTC | 2 |
| 316 | 2021 | 267 | 2021-05-14 12:00:00 UTC | 2 |
| 317 | 2021 | 272 | 2021-05-17 00:00:00 UTC | 3 |
| 318 | 2021 | 273 | 2021-05-17 12:00:00 UTC | 2 |
| 319 | 2021 | 278 | 2021-05-20 00:00:00 UTC | 2 |
| 320 | 2021 | 280 | 2021-05-21 00:00:00 UTC | 4 |
| 321 | 2021 | 281 | 2021-05-21 12:00:00 UTC | 3 |
| 322 | 2021 | 285 | 2021-05-23 12:00:00 UTC | 4 |
| 323 | 2021 | 288 | 2021-05-25 00:00:00 UTC | 4 |
| 324 | 2021 | 291 | 2021-05-26 12:00:00 UTC | 3 |
| 325 | 2021 | 293 | 2021-05-27 12:00:00 UTC | 3 |
| 326 | 2021 | 298 | 2021-05-30 00:00:00 UTC | 2 |
| 327 | 2021 | 301 | 2021-05-31 12:00:00 UTC | 3 |
| 328 | 2021 | 302 | 2021-06-01 00:00:00 UTC | 4 |
| 329 | 2021 | 303 | 2021-06-01 12:00:00 UTC | 2 |
| 330 | 2021 | 305 | 2021-06-02 12:00:00 UTC | 3 |
| 331 | 2021 | 308 | 2021-06-04 00:00:00 UTC | 2 |
| 332 | 2021 | 309 | 2021-06-04 12:00:00 UTC | 2 |
| 333 | 2021 | 311 | 2021-06-05 12:00:00 UTC | 3 |
| 334 | 2021 | 317 | 2021-06-08 12:00:00 UTC | 2 |
| 335 | 2021 | 326 | 2021-06-13 00:00:00 UTC | 3 |
| 336 | 2021 | 334 | 2021-06-17 00:00:00 UTC | 2 |
| 337 | 2021 | 357 | 2021-06-28 12:00:00 UTC | 2 |
| 338 | 2021 | 464 | 2021-08-21 00:00:00 UTC | 2 |
| 339 | 2021 | 466 | 2021-08-22 00:00:00 UTC | 2 |
| 340 | 2021 | 470 | 2021-08-24 00:00:00 UTC | 2 |
| 341 | 2021 | 471 | 2021-08-24 12:00:00 UTC | 3 |
| 342 | 2021 | 481 | 2021-08-29 12:00:00 UTC | 2 |
| 343 | 2021 | 482 | 2021-08-30 00:00:00 UTC | 3 |
| 344 | 2021 | 500 | 2021-09-08 00:00:00 UTC | 3 |
| 345 | 2021 | 503 | 2021-09-09 12:00:00 UTC | 5 |
| 346 | 2021 | 507 | 2021-09-11 12:00:00 UTC | 4 |
| 347 | 2021 | 511 | 2021-09-13 12:00:00 UTC | 2 |
| 348 | 2021 | 512 | 2021-09-14 00:00:00 UTC | 4 |
| 349 | 2021 | 513 | 2021-09-14 12:00:00 UTC | 5 |
| 350 | 2021 | 517 | 2021-09-16 12:00:00 UTC | 4 |
| 351 | 2021 | 519 | 2021-09-17 12:00:00 UTC | 3 |
| 352 | 2021 | 521 | 2021-09-18 12:00:00 UTC | 2 |
| 353 | 2021 | 522 | 2021-09-19 00:00:00 UTC | 4 |
| 354 | 2021 | 523 | 2021-09-19 12:00:00 UTC | 3 |
| 355 | 2021 | 524 | 2021-09-20 00:00:00 UTC | 3 |
| 356 | 2021 | 525 | 2021-09-20 12:00:00 UTC | 3 |
| 357 | 2021 | 526 | 2021-09-21 00:00:00 UTC | 2 |
| 358 | 2021 | 527 | 2021-09-21 12:00:00 UTC | 2 |
| 359 | 2021 | 531 | 2021-09-23 12:00:00 UTC | 2 |
| 360 | 2021 | 532 | 2021-09-24 00:00:00 UTC | 3 |
| 361 | 2021 | 533 | 2021-09-24 12:00:00 UTC | 3 |
| 362 | 2021 | 535 | 2021-09-25 12:00:00 UTC | 2 |
| 363 | 2021 | 536 | 2021-09-26 00:00:00 UTC | 2 |
| 364 | 2021 | 541 | 2021-09-28 12:00:00 UTC | 2 |
| 365 | 2021 | 546 | 2021-10-01 00:00:00 UTC | 5 |
| 366 | 2021 | 547 | 2021-10-01 12:00:00 UTC | 4 |
| 367 | 2021 | 550 | 2021-10-03 00:00:00 UTC | 3 |
| 368 | 2021 | 551 | 2021-10-03 12:00:00 UTC | 3 |
| 369 | 2021 | 556 | 2021-10-06 00:00:00 UTC | 4 |
| 370 | 2021 | 566 | 2021-10-11 00:00:00 UTC | 3 |
| 371 | 2021 | 577 | 2021-10-16 12:00:00 UTC | 5 |
| 372 | 2021 | 587 | 2021-10-21 12:00:00 UTC | 2 |
| 373 | 2021 | 599 | 2021-10-27 12:00:00 UTC | 4 |
| 374 | 2021 | 601 | 2021-10-28 12:00:00 UTC | 3 |
| 375 | 2021 | 607 | 2021-10-31 12:00:00 UTC | 2 |
| 376 | 2021 | 612 | 2021-11-03 00:00:00 UTC | 4 |
| 377 | 2021 | 613 | 2021-11-03 12:00:00 UTC | 3 |
| 378 | 2021 | 615 | 2021-11-04 12:00:00 UTC | 2 |
| 379 | 2021 | 616 | 2021-11-05 00:00:00 UTC | 2 |
| 380 | 2021 | 619 | 2021-11-06 12:00:00 UTC | 2 |
| 381 | 2021 | 634 | 2021-11-14 00:00:00 UTC | 2 |
| 382 | 2021 | 636 | 2021-11-15 00:00:00 UTC | 2 |
| 383 | 2021 | 639 | 2021-11-16 12:00:00 UTC | 2 |
| 384 | 2021 | 653 | 2021-11-23 12:00:00 UTC | 2 |
| 385 | 2021 | 656 | 2021-11-25 00:00:00 UTC | 3 |
| 386 | 2021 | 663 | 2021-11-28 12:00:00 UTC | 2 |
| 387 | 2021 | 668 | 2021-12-01 00:00:00 UTC | 2 |
| 388 | 2021 | 672 | 2021-12-03 00:00:00 UTC | 2 |
| 389 | 2021 | 675 | 2021-12-04 12:00:00 UTC | 4 |
| 390 | 2021 | 678 | 2021-12-06 00:00:00 UTC | 3 |
| 391 | 2021 | 681 | 2021-12-07 12:00:00 UTC | 3 |
| 392 | 2021 | 685 | 2021-12-09 12:00:00 UTC | 5 |
| 393 | 2021 | 696 | 2021-12-15 00:00:00 UTC | 2 |
| 394 | 2021 | 698 | 2021-12-16 00:00:00 UTC | 4 |
| 395 | 2021 | 699 | 2021-12-16 12:00:00 UTC | 2 |
| 396 | 2021 | 701 | 2021-12-17 12:00:00 UTC | 2 |
| 397 | 2021 | 703 | 2021-12-18 12:00:00 UTC | 3 |
| 398 | 2021 | 709 | 2021-12-21 12:00:00 UTC | 2 |
| 399 | 2021 | 711 | 2021-12-22 12:00:00 UTC | 3 |
| 400 | 2021 | 712 | 2021-12-23 00:00:00 UTC | 4 |
| 401 | 2021 | 713 | 2021-12-23 12:00:00 UTC | 3 |
| 402 | 2021 | 714 | 2021-12-24 00:00:00 UTC | 6 |
| 403 | 2021 | 716 | 2021-12-25 00:00:00 UTC | 2 |
| 404 | 2021 | 717 | 2021-12-25 12:00:00 UTC | 5 |
| 405 | 2021 | 719 | 2021-12-26 12:00:00 UTC | 5 |
| 406 | 2021 | 722 | 2021-12-28 00:00:00 UTC | 4 |
| 407 | 2021 | 723 | 2021-12-28 12:00:00 UTC | 3 |
| 408 | 2021 | 724 | 2021-12-29 00:00:00 UTC | 6 |
| 409 | 2021 | 725 | 2021-12-29 12:00:00 UTC | 4 |
| 410 | 2021 | 726 | 2021-12-30 00:00:00 UTC | 6 |
| 411 | 2022 | 0 | 2022-01-01 00:00:00 UTC | 3 |
| 412 | 2022 | 6 | 2022-01-04 00:00:00 UTC | 7 |
| 413 | 2022 | 9 | 2022-01-05 12:00:00 UTC | 3 |
| 414 | 2022 | 10 | 2022-01-06 00:00:00 UTC | 3 |
| 415 | 2022 | 13 | 2022-01-07 12:00:00 UTC | 2 |
| 416 | 2022 | 14 | 2022-01-08 00:00:00 UTC | 3 |
| 417 | 2022 | 15 | 2022-01-08 12:00:00 UTC | 3 |
| 418 | 2022 | 16 | 2022-01-09 00:00:00 UTC | 4 |
| 419 | 2022 | 19 | 2022-01-10 12:00:00 UTC | 4 |
| 420 | 2022 | 20 | 2022-01-11 00:00:00 UTC | 4 |
| 421 | 2022 | 21 | 2022-01-11 12:00:00 UTC | 4 |
| 422 | 2022 | 22 | 2022-01-12 00:00:00 UTC | 4 |
| 423 | 2022 | 23 | 2022-01-12 12:00:00 UTC | 4 |
| 424 | 2022 | 24 | 2022-01-13 00:00:00 UTC | 8 |
| 425 | 2022 | 25 | 2022-01-13 12:00:00 UTC | 4 |
| 426 | 2022 | 26 | 2022-01-14 00:00:00 UTC | 4 |
| 427 | 2022 | 27 | 2022-01-14 12:00:00 UTC | 9 |
| 428 | 2022 | 28 | 2022-01-15 00:00:00 UTC | 4 |
| 429 | 2022 | 29 | 2022-01-15 12:00:00 UTC | 2 |
| 430 | 2022 | 30 | 2022-01-16 00:00:00 UTC | 5 |
| 431 | 2022 | 31 | 2022-01-16 12:00:00 UTC | 2 |
| 432 | 2022 | 32 | 2022-01-17 00:00:00 UTC | 3 |
| 433 | 2022 | 33 | 2022-01-17 12:00:00 UTC | 7 |
| 434 | 2022 | 34 | 2022-01-18 00:00:00 UTC | 4 |
| 435 | 2022 | 35 | 2022-01-18 12:00:00 UTC | 8 |
| 436 | 2022 | 36 | 2022-01-19 00:00:00 UTC | 7 |
| 437 | 2022 | 37 | 2022-01-19 12:00:00 UTC | 8 |
| 438 | 2022 | 38 | 2022-01-20 00:00:00 UTC | 3 |
| 439 | 2022 | 39 | 2022-01-20 12:00:00 UTC | 4 |
| 440 | 2022 | 41 | 2022-01-21 12:00:00 UTC | 6 |
| 441 | 2022 | 43 | 2022-01-22 12:00:00 UTC | 6 |
| 442 | 2022 | 44 | 2022-01-23 00:00:00 UTC | 5 |
| 443 | 2022 | 46 | 2022-01-24 00:00:00 UTC | 4 |
| 444 | 2022 | 47 | 2022-01-24 12:00:00 UTC | 3 |
| 445 | 2022 | 48 | 2022-01-25 00:00:00 UTC | 3 |
| 446 | 2022 | 50 | 2022-01-26 00:00:00 UTC | 2 |
| 447 | 2022 | 52 | 2022-01-27 00:00:00 UTC | 3 |
| 448 | 2022 | 53 | 2022-01-27 12:00:00 UTC | 6 |
| 449 | 2022 | 54 | 2022-01-28 00:00:00 UTC | 4 |
| 450 | 2022 | 55 | 2022-01-28 12:00:00 UTC | 3 |
| 451 | 2022 | 60 | 2022-01-31 00:00:00 UTC | 3 |
| 452 | 2022 | 61 | 2022-01-31 12:00:00 UTC | 4 |
| 453 | 2022 | 62 | 2022-02-01 00:00:00 UTC | 3 |
| 454 | 2022 | 63 | 2022-02-01 12:00:00 UTC | 2 |
| 455 | 2022 | 64 | 2022-02-02 00:00:00 UTC | 4 |
| 456 | 2022 | 74 | 2022-02-07 00:00:00 UTC | 3 |
| 457 | 2022 | 79 | 2022-02-09 12:00:00 UTC | 3 |
| 458 | 2022 | 135 | 2022-03-09 12:00:00 UTC | 2 |
| 459 | 2022 | 140 | 2022-03-12 00:00:00 UTC | 2 |
| 460 | 2023 | 719 | 2023-12-26 12:00:00 UTC | 2 |

**Statements of Assumptions**

*Notation*

Let $\mathrm{P}(X|V)$ denote a generic probability statement, $f_X(X|V)$ represent a general probability density (or mass) function, used when $X$ is continuous (or mixed).

Let racial and ethnic group $G$ serve as a proxy for skin tone, where $G = 1$ denotes the disadvantaged group (e.g., non-Hispanic Black [NHB]) and $G = 0$ the advantaged group (e.g., non-Hispanic White [NHW]). Let $T \in (0,1)$ indicate membership in the standard population, where $T = 1$ denotes inclusion in the chosen standard population.

Let $Z$ represent the exposure of interest ($SpO_2$) and $\tilde{Z}$ denote the set of intervention strategies. Each $\tilde{z}$ is characterized by a PO with a designated bias profile that determines each group's distribution of $Z$, i.e., $f_{\tilde{z}}(Z|\cdot)$, taking values continuously within a clinically relevant range (e.g., 70-100%).

Let $Y_{k+J}$ be the observed outcome of interest (e.g., dexamethasone treatment), $\boldsymbol{A_k}$ denote the set of allowable covariates (e.g., clinical need), and $\boldsymbol{N_k}$ denote the set of non-allowable covariates (e.g., SES), all indexed by calendar time $k$. Let $Y_{k+J}^{\tilde{z}}$ be the potential outcome measured at time $k + J$ under the hypothetical intervention strategy $\tilde{z}$.

1) Overlap: At each calendar time $k$, persons from each social group $G = g$ (e.g., NHB and NHW) are observed at each level of the allowable covariates $\boldsymbol{A_k}$ (e.g., clinical need) observed within the standard population ($T = 1$). This ensures that $\boldsymbol{A_k}$ can be balanced across groups so that meaningful contrasts of disparity can be defined.
   Formally,
   $P(G = g|\, \boldsymbol{A_k} = \boldsymbol{a}, k) > 0$ for all $a$ with $f_{\boldsymbol{A_k}}(\boldsymbol{A_k}|T = 1, k) > 0$ and all $g \in G$

2) Innocuous Sampling: At each calendar time $k$, among eligible individuals within a given social group $G = g$, sampled and unsampled individuals have the same conditional distribution of the outcome $Y_{k+J}$ given the covariates $(\boldsymbol{N_k}, \boldsymbol{A_k})$ and $Z$.
   $$P^{sampled}\left(Y_{k+J} = y \middle| G = g, Z, \boldsymbol{N_k}, \boldsymbol{A_k}, k\right) = P^{unsampled}\left(Y_{k+J} = y \middle| G = g, Z, \boldsymbol{N_k}, \boldsymbol{A_k}, k\right)$$

3) Conditional Exchangeability/No Unmeasured Confounding: The potential outcome under intervention strategy $\tilde{z} \in (\tilde{z}_0, \tilde{z}_1, \tilde{z}_2, \tilde{z}_3, \tilde{z}_4, \tilde{z}_5, \tilde{z}_6, \tilde{z}_7, \tilde{z}_8)$, denoted $Y_{k+J}^{\tilde{z}}$, is conditionally independent of the observed exposure $Z$, given social group $G = g$, allowable covariates $\boldsymbol{A_k}$, and non-allowable covariates $\boldsymbol{N_k}$. This implies that all relevant confounders of the relationship between $SpO_2$ and outcome are measured and adjusted for.
   Formally,

$Y_{k+J}^{\tilde{z}} \perp Z | G = g, \boldsymbol{N_k}, \boldsymbol{A_k}, k$ for all $z$ within the support of each interventional strategy $f_{\tilde{z}}(Z | \cdot)$.

4) Positivity: Within each social group $G = g$, for every combination of allowable $\boldsymbol{A_k}$ and non-allowable $\boldsymbol{N_k}$ covariates, there must be a positive probability of observing an $SpO_2$ value consistent with the intervention strategy $\tilde{z}$, i.e., within the range of $Z$ implied by $\tilde{z}$. This ensures that the effect of $\tilde{z}$ is identifiable within all relevant strata. In our setting, this means that for each $(\boldsymbol{a_k}, \boldsymbol{n_k})$ profile, it is theoretically possible to observe $SpO_2$ values that would occur if patients were assigned to devices with the bias level specified by $\tilde{z}$. In other words, the conditional density of $SpO_2$ values must be positive within each stratum over the range implied by the intervention strategy $\tilde{z}$.
Formally,
$f_z(z | G = g, \boldsymbol{A_k}, \boldsymbol{N_k}, k) > 0$ for all $(\boldsymbol{N_k}, \boldsymbol{A_k})$ with $f_{\boldsymbol{A_k}, \boldsymbol{N_k}}(\boldsymbol{A_k}, \boldsymbol{N_k} | G = g, k) > 0$ and all z within the support implied by $\tilde{Z} = \tilde{z}$, i.e., $f_{\tilde{z}}(Z | \cdot)$.

   Although $SpO_2$ values under some combinations of $(\boldsymbol{a_k}, \boldsymbol{n_k})$ may be sparsely represented in the data, such practical (random) positivity violations are attributed to finite sample size and do not preclude causal identification under the assumption that the assignment mechanism remains theoretically possible.[21]

5) Consistency: The potential outcome under intervention $\tilde{z}$ equals the observed outcome when a person receives device $\tilde{z}$. In our setting, this implies that the potential outcome under an $SpO_2$ value $z$ induced by being assigned to device $\tilde{z}$ is the same as if that value were observed in practice. This requires that the intervention be well-defined and a person's outcome depends only on their realized $SpO_2$ value, not on the specific mechanism by which that value arises. It also requires no interference: each person's outcome depends only on their own interventional $SpO_2$ assignment and not on the assignments of others. We consider these conditions are plausible because dexamethasone treatment decisions are guided by a patient's observed oxygen saturation, and dexamethasone was not a limited resource during the study period, reducing the likelihood that one patient's treatment would affect another's access.
Formally,
If $\tilde{z} \Rightarrow Z$ then $Y_{k+J}^{\tilde{z}} \equiv Y_{k+J}$

**Proofs of Identification Formulas**

Recall $Z$ is continuous, and $\boldsymbol{N_k}$ and $\boldsymbol{A_k}$ are covariates that may include both continuous (e.g., age) and discrete (e.g., sex) components. We express the estimand in integral form as:

$$\mu_k^{\tilde{z}}(g) = \int E\left[Y_{k+J}^{\tilde{z}} \middle| G = g, \boldsymbol{A_k} = \boldsymbol{a}, k\right] f_{A_k}(\boldsymbol{A_k} = \boldsymbol{a} | T = 1, k) da$$

Note that the estimand assumes that the innocuous sampling and overlap assumptions hold.

The identification formula is:

$$\mu_k^{\tilde{z}}(g) = \int \mathrm{E}\left[Y_{k+J}^{\tilde{z}} \middle| G = g, \boldsymbol{A_k} = \boldsymbol{a}, k\right] f_{\boldsymbol{A_k}}(\boldsymbol{A_k} = \boldsymbol{a} | T = 1, k) da$$
$$= \int \int E\left[Y_{k+J}^{\tilde{z}} \middle| G = g, \boldsymbol{N_k} = \boldsymbol{n}, \boldsymbol{A_k} = \boldsymbol{a}, k\right] \times f_{\boldsymbol{N_k}}(\boldsymbol{N_k} = \boldsymbol{n} | G = g, \boldsymbol{A_k} = \boldsymbol{a}, k) dn$$
$$\times f_{\boldsymbol{A_k}}(\boldsymbol{A_k} = \boldsymbol{a} | T = 1, k) da$$
$$= \int \int \int \mathrm{E}\left[Y_{k+J}^{\tilde{z}} \middle| G = g, Z = z, \boldsymbol{N_k} = \boldsymbol{n}, \boldsymbol{A_k} = \boldsymbol{a}, k\right]$$
$$\times f_{\tilde{z}}(Z = z | G = g, \boldsymbol{N_k} = \boldsymbol{n}, \boldsymbol{A_k} = \boldsymbol{a}, k) dz \times f_{\boldsymbol{N_k}}(\boldsymbol{N_k} = \boldsymbol{n} | G = g, \boldsymbol{A_k} = \boldsymbol{a}, k) dn$$
$$\times f_{\boldsymbol{A_k}}(\boldsymbol{A_k} = \boldsymbol{a} | T = 1, k) da$$
$$= \int \int \int \mathrm{E}\left[Y_{k+J} \middle| G = g, Z = z, \boldsymbol{N_k} = \boldsymbol{n}, \boldsymbol{A_k}, k\right] f_{\tilde{z}}(Z = z | G = g, \boldsymbol{N_k} = \boldsymbol{n}, \boldsymbol{A_k} = \boldsymbol{a}, k) dz$$
$$\times f_{\boldsymbol{N_k}}(\boldsymbol{N_k} = \boldsymbol{n} | G = g, \boldsymbol{A_k} = \boldsymbol{a}, k) dn \times f_{\boldsymbol{A_k}}(\boldsymbol{A_k} = \boldsymbol{a} | T = 1, k) da$$

where $f_{\tilde{z}}(Z | G = g, \boldsymbol{N_k}, \boldsymbol{A_k}, k)$ represents an interventional distribution of $SpO_2$ according to the bias parameters of the assigned PO device.

The first line of the formula holds by the overlap assumption, the first equality holds by the total law of probability, the second by conditional exchangeability and positivity, and the third by consistency.

If all variables are discrete, we would just follow the same logic using summation.

Estimand:

$$\mu_k^{\tilde{z}}(g) = \sum_{a,k} E\left[Y_{k+J}^{\tilde{z}} \middle| G = g, \boldsymbol{A_k} = \boldsymbol{a}, k\right] P\left(\boldsymbol{A_k} = \boldsymbol{a} | T = 1, k\right)$$

Identifying formula:

$$\sum_{a,k} E\left[Y_{k+J}^{\tilde{z}} \middle| G = g, \boldsymbol{A_k} = \boldsymbol{a}, k\right] P(\boldsymbol{A_k} = \boldsymbol{a} | T = 1, k)$$
$$= \sum_{a,k,n_k} E\left[Y_{k+J}^{\tilde{z}} \middle| G = g, \boldsymbol{N_k} = \boldsymbol{n}, \boldsymbol{A_k} = \boldsymbol{a}, k\right] P(\boldsymbol{N_k} = \boldsymbol{n} | G = g, \boldsymbol{A_k} = \boldsymbol{a}, k) P(\boldsymbol{A_k} = \boldsymbol{a} | T = 1, k)$$
$$= \sum_{a,k,n,z} E\left[Y_{k+J}^{\tilde{z}} \middle| G = g, Z = z, \boldsymbol{N_k} = \boldsymbol{n}, \boldsymbol{A_k} = \boldsymbol{a}, k\right] P_{\tilde{z}}(Z = z | G = g, \boldsymbol{N_k} = \boldsymbol{n}, \boldsymbol{A_k} = \boldsymbol{a}, k) P(\boldsymbol{N_k} = \boldsymbol{n} | G = g, \boldsymbol{A_k} = \boldsymbol{a}, k) P(\boldsymbol{A_k} = \boldsymbol{a} | T = 1, k)$$
$$= \sum_{a,k,n,z} E\left[Y_{k+J} \middle| G = g, Z = z, \boldsymbol{N_k} = \boldsymbol{n}, \boldsymbol{A_k} = \boldsymbol{a}, k\right] P_{\tilde{z}}(Z = z | G = g, \boldsymbol{N_k} = \boldsymbol{n}, \boldsymbol{A_k} = \boldsymbol{a}, k) P(\boldsymbol{N_k} = \boldsymbol{n} | G = g, \boldsymbol{A_k} = \boldsymbol{a}, k) P(\boldsymbol{A_k} = \boldsymbol{a} | T = 1, k)$$

**Semiparametric G-computation Steps for Time-to-Event Outcomes**

We extend the semiparametric G-computation procedure[22,23] to time-to-event outcomes. The data structure for binary outcomes is described in the main paper (Section 4.2). Here we introduce additional notation to accommodate time-to-event outcomes.

$J_{i,k}$: time to treatment for person $i$ enrolled at calendar time $k$.

$C_{i,k}$: censoring time.

$J_{i,k}^* = \min(J_{i,k}, C_{i,k})$, either the treatment time or censoring time, whichever occurs first.

$\delta_{i,k} = I(J_{i,k}^* \leq C_{i,k})$: event indicator, 1 if the treatment was observed before censoring and 0 otherwise.

We estimate the standardized survival function $S^{\tilde{z}}(j) = P(J_k^{\tilde{z}} > j)$, i.e., the cumulative probability of remaining untreated by time $j$ under a hypothetical intervention $\tilde{z}$. Adapting from Edwards et al.,[22] the procedure proceeds as follows.

Step 1: Estimate the hazard function and the cumulative baseline hazard function

Within each social group $G = g$, we fit a Cox proportional hazard model on observed $SpO_2$ value $Z$, allowables $\boldsymbol{A_k}$, non-allowables $\boldsymbol{N_k}$, and calendar time $k$.

The hazard function is specified as:

$$\lambda(j|G = g) = \lambda_0(j|G = g)e^{\boldsymbol{\beta_g} h(\boldsymbol{W})}$$

where $h(\boldsymbol{W})$ is a function of covariates $\boldsymbol{W} = \{Z, \boldsymbol{A_k}, \boldsymbol{N_k}, k\}$, $\boldsymbol{\beta_g}$ is the group-specific vector of coefficients for $\boldsymbol{W}$.

The Breslow estimator of the cumulative baseline hazard function among $G = g$ is:

$$\widehat{\Lambda}_0(j|G = g) = \sum_{i=0}^{n} \frac{I(G = g)I(J_{i,k}^* \leq j)\delta_{i,k}}{\sum_{m=1}^{n} I(G = g)R_m(J_{i,k}^*)\, exp[\widehat{\boldsymbol{\beta}}_{\boldsymbol{g}} h(\boldsymbol{W_m})]}$$

where $R_m(J_{i,k}^*)$ is an indicator that person $m$ is in the risk set at time $J_{i,k}^*$, and $\widehat{\boldsymbol{\beta}}_{\boldsymbol{g}}$ is estimated from the Cox model fit among those with $G = g$ adjusting for covariates $\boldsymbol{W}$.

Step 2: Assign intervention and predict survival

For a given intervention $\tilde{z}$, we replace the observed $SpO_2$ value $Z$ with an intervened value $Z^*$, yielding $\boldsymbol{W}^* = \{Z = Z^*, \boldsymbol{A_k}, \boldsymbol{N_k}, k\}$. Using the Breslow estimator from Step 1, we predict the individual-level survival function at each event time $j$:

$$\hat{S}(j|G = g, \boldsymbol{W}^*) = exp\{-\widehat{\Lambda}_0(j|G = g)\, exp[\widehat{\boldsymbol{\beta}}_{\boldsymbol{g}} h(\boldsymbol{W}^*)]\}.$$

Steps 3-5: Standardization over allowable covariates and calendar time

At each event time $j$, we standardize the predicted survival function by averaging over the joint empirical distribution of allowable covariates $\boldsymbol{A_k}$ and calendar time $k$ in the standard population. This yields a group-specific standardized survival curve $\hat{S}_g^{\tilde{z}}(j)$.

We summarize time to treatment by estimating the restricted mean survival time (RMST)[24] up to 24 hours as

$$RMST_g^{\tilde{z}} = \int_0^{24} \hat{S}_g^{\tilde{z}}(j)\, dj,$$

which we compute numerically using the trapezoidal rule. RMST represents the average time spent untreated within the 24-hour window under intervention $\tilde{z}$.

**Aggregation of Results Across Calendar Time**

Recall our TS+TT study is conceptualized as a series of nested trials indexed by calendar time $k$. This design avoids assuming that disparity or intervention effects are constant over time, allowing us to capture clinically relevant variation in treatment patterns and provider responses.[25] To facilitate interpretation, we aggregate results across calendar time (e.g., akin to a meta-analysis) by practically treating $k$ as an allowable covariate and standardizing to its distribution in the standard population $T = 1$ pooled across calendar times, similar to the procedure discussed in detail in Jackson et al. (2025).[25] This yields an aggregated estimate interpretable as the average intervention effect on disparity across calendar time while addressing potential bias from differential enrollment timing across social groups.[25] Operationally, this is implemented by integrating over $f_{\boldsymbol{A_k},k}(\boldsymbol{A_k}, k|T = 1)$ in the estimand definition and is reflected in our G-computation steps described in the main paper.

There are multiple ways to code the enrollment windows. One straightforward approach is to designate the earliest possible window as k=0 and index subsequent windows in 12-hour increments within each calendar year. For example, in 2020, the earliest window (2020-01-01 at 00:00) is coded as k=0; the next window, k=1, corresponds to 2020-01-01 at 12:00; k=2 to 2020-01-02 at 00:00; and so forth. The same coding scheme is used for other years. For a 365-day year, this yields 730 possible enrollment windows (732 in a leap year). In the analysis, we include an indicator for calendar year and model k using a natural spline to flexibly account for temporal trends in disparities or intervention effects. This coding scheme allows for year and seasonal effects but assumes the absence of year-specific seasonal effects. This assumption could be relaxed by using separate spline specifications for each calendar year.

**Illustration of Stratified Sampling to Balance Allowable Covariates Across Social Groups**

| Stratum | NHB sampling frame (N=800) | Intended sampling fraction | Sample | NHW sampling frame (N=600) | Intended sampling fraction | Rescaled sampling fraction | Sample |
|---|---|---|---|---|---|---|---|
| <65 yrs, male | 100 | 1 | 100 | 100 | 1 | 0.5 | 50 |
| <65 yrs, female | 100 | 1 | 100 | 100 | 1 | 0.5 | 50 |
| ≥65 yrs, male | 200 | 1 | 200 | 200 | 1 | 0.5 | 100 |
| ≥65 yrs, female | 400 | 1 | 400 | 200 | 2 | 1 | 200 |

The stratified sampling procedure is formally presented in Jackson et al. (2025).[25] To intuit the stratified sampling procedure, consider a simplified example in which allowable covariates consist of age (<65 vs. ≥65 years) and sex (male vs. female), yielding four strata. Suppose that the distribution of these covariates differs between the NHB and NHW source populations.

In this example, we designate NHB as the standard population. As a result, all eligible NHB individuals are selected, and the intended sampling fraction for each NHB stratum is 1. The goal is then to sample from the NHW population in a way that reproduces the same age-sex distribution observed in the NHB population. For each stratum, we compute an intended sampling fraction for NHW that reflects how common that stratum is in the NHB population relative to the NHW population. In the first three strata, this fraction equals 1, indicating that all NHW individuals in those strata would be selected. However, in the ≥65 yrs, female stratum, the intended sampling fraction exceeds 1, implying that more individuals would need to be sampled than are available.

Because sampling without replacement cannot exceed the available population, we rescale the sampling fractions across all strata by a common factor. This preserves the relative sampling weights across strata while ensuring that all rescaled sampling fractions lie between 0 and 1. After rescaling, individuals in each stratum are randomly sampled according to the adjusted sampling fractions, which are now probabilities of inclusion.

This procedure yields an analytic sample in which a person only appears once per trial and the distribution of allowable covariates in the NHW group matches that of the NHB standard population.

**Estimation of the Natural Course for Model Validation**

Our G-computation procedure involves two distinct modeling components: (a) an outcome model used to generate counterfactual predictions for each individual based on their covariate profile, and (b) a standardization model that aggregates these predictions to a common distribution of allowable covariates. We therefore assessed potential model misspecification at both stages by estimating outcomes under the natural course.

(a) Natural course for outcome model validation

We first assessed the specification of the outcome model by estimating outcomes under the natural course prior to standardization. Estimation of the natural course requires modeling the observed data-generating process under observed exposure rather than setting the exposure under a specified intervention. In our setting, we modeled the distribution of measurement error conditional on covariates for each social group. For each individual, we generated error values under the natural course by drawing from a normal distribution centered at the model-predicted mean error for that individual's covariate pattern, with variance given by the estimated residual variance. We then added this simulated error to the individual's observed $SaO_2$ to construct the corresponding $SpO_2$ value under the natural course. Using these simulated $SpO_2$ values, we applied the group-specific outcome model to obtain predicted outcomes.

We compared the resulting model-predicted outcomes under the natural course with the observed outcomes in the data. Agreement between these quantities provides a diagnostic check on the adequacy of the outcome model specification.

(b) Natural course for standardization model validation

We next evaluated the specification of the standardization model, which aggregates individual-level predictions over the distribution of allowable covariates and calendar time. To do so, we applied the standardization step to the outcome predictions obtained above, yielding standardized outcomes under the natural course.

We then compared these standardized natural course estimates with the corresponding empirically standardized outcomes computed from the observed data. Provided that the outcome model passes the natural course diagnostic, agreement between these quantities provides a diagnostic check on the adequacy of the standardization model specification.

Together, these two natural course comparisons allow us to separately probe potential misspecification of the outcome model and the standardization model within the G-computation algorithm. As shown in **sTable 5**, natural course estimates closely aligned with the corresponding observed quantities both before and after standardization for the binary outcome of 24-hour treatment receipt. Small differences between observed and natural course RMST estimates suggest modest residual misspecification, which may arise from time-specific standardization and the accumulation of estimation error across event times, but the estimates were imprecise.

Overall, these results support adequate specification of both the outcome and standardization models.

## Missing Data Handling and Multiple Imputation

*Overview of Missingness*

The enrolled eligible population consisted of 10,450 observations. As shown in **sTable 6**, 6 covariates had complete data, while 10 covariates exhibited varying degrees of missingness. Missingness was most substantial for hemoglobin (29.8%) and ADI (28.9%), followed by BMI (20.9%), smoking status (8.8%), bilirubin (8.6%), and SBP (7.2%). Pulse rate (1.4%), respiratory rate (1.7%), body temperature (1.6%), and creatinine (0.05%) had low levels of missingness.

*Pre-processing to Reduce Structural Missingness*

Before imputation, we applied targeted preprocessing steps to reduce avoidable structural missingness while preserving interpretability and alignment with the data-generating process.

<u>BMI:</u> For individuals with repeated enrollment windows, BMI values were carried forward and backward within-person, assuming minimal short-term variation. This reduced missingness from approximately 21% to 4%, after which remaining values were imputed.

<u>Smoking status:</u> Smoking status was carried forward and backward within-person when available at any time, assuming relative stability over the study period. This reduced missingness to approximately 4%, after which remaining values were imputed.

<u>ADI:</u> ADI missingness primarily reflected unavailable block group information, with a small proportion due to suppressed values. We imputed ADI national percentile (0-100) and included an indicator for suppressed values. We did not attempt to impute block group directly due to high dimensionality and concerns about unstable estimation. Missing ADI may plausibly reflect mechanisms such as out-of-state residence or incomplete address capture.

<u>Laboratory measures (hemoglobin, bilirubin, creatinine):</u> Missingness in laboratory measures was largely attributable to lack of ordering (i.e., unobserved). We note that for our treatment-based outcome, it is the results from ordered labs that are causally efficacious, as these are known to the healthcare provider at the time of treatment decision-making. The healthcare provider does not have access to results of unordered labs. To reflect this measurement process, which hinges on lab orders, we included an indicator for unordered lab status and imputed missing lab values among these unordered labs using the mean among observed lab result values, representing "not measured" rather than latent disease severity.

*Multiple Imputation Model Specification*

Missing covariate data were imputed using multiple imputation by chained equations (MICE), implemented via the mice package in R. We generated 5 imputed datasets using 10 iterations per dataset. We implemented a two-level (multilevel) multiple imputation framework to account for repeated enrollment windows nested within individuals.

Clustering structure: A person-level identifier was specified as the level-2 clustering variable and excluded as a predictor except for defining clusters.

Variables included: The imputation model included all variables used in the outcome analysis (allowable covariates $A_k$, non-allowable covariates $N_k$, calendar time $k$, and outcomes), as well as auxiliary variables plausibly related to missingness, including hospital location, oxygen use at the time of $SpO_2$ measurement, ICU status, vasopressor use in the prior enrollment window, and supplemental oxygen use in the prior 24 hours.

Imputation methods:

Level 1 (within-person, time-varying): pulse rate, respiratory rate, body temperature, and SBP were imputed using multilevel linear models.

Level 2 (person-level): BMI, smoking status, and ADI percentile were imputed using multilevel predictive mean matching.

Diagnostic Checking of Imputation:

We assessed convergence and adequacy of the imputation models using standard graphical diagnostics. Trace plots of imputed values across iterations showed stable mixing without evidence of non-convergence. Strip plots comparing observed and imputed values demonstrated substantial overlap and no systematic deviations, suggesting that the imputed values were plausible and consistent with the observed data distributions.

### Relation to Other Estimands

Here, we very briefly distinguish how our estimand as expressed in Equation (2) differs from others that involve quantifying counterfactual disparities under hypothetical interventions.

When applied in the disparities context, causal mediation analysis often frames a total effect of social group membership as a form of disparity[26,27] and focuses on the effects of changing social group membership while simultaneously holding fixed variables potentially affected by social group membership to a specific value (controlled direct effects) or to "natural values" arising under the absence of treatment (natural direct/indirect effects).[28,29] The latter are used to investigate path-specific effects by which interventions to change social group membership or its perception by some decision-maker affect the outcome. Mediation analysis has also been conceptualized in other ways,[30] such as organic effects,[31] analogues of NDEs and path-specific effects,[32,33] separable effects,[34] and most recently, information-based effects.[35] Regardless of their philosophical differences, each of these mediation estimands envisions an intervention to change who belongs to which social group or an intervention to change the perception of who belongs to which social group. This differs from the estimand reflected by our TS+TT design, where there is no intervention to change social group membership or its perception, only an intervention to impact a determinant of the outcome. The estimand reflected by our TS+TT design is not a causal mediation analysis and does not attempt to express or even approximate direct or indirect effects of social group membership though intermediates.

A second class of estimands are forms of population impact,[36,37] wherein social groups are not intervened upon, but their outcome distributions are compared upon a hypothetical intervention to change the distribution of some determinant of the outcome. Earlier literature that has proposed these estimands framed them as being related to causal mediation analysis where social group membership is intervened upon as the exposure, perhaps because of their similar identifying formulae obtained under weaker identifying assumptions.[38,39] However, it has been shown that such framing is unwarranted, and also unnecessary, as such estimands can be non-null even in the absence of mediation.[40] More recent literature has emphasized the population impact interpretation,[41] focusing on how comparative outcome distributions change under intervention, which is inherently a policy-relevant focus. As pointed out in our introduction, these estimands treat all factors as non-allowable, and thus are only substantively relevant for cases where pure inequality can be taken as a measure of disparity. Our TS+TT design, through its initial stratified sampling step, overcomes this limitation. It does so by first enrolling a sample where social groups are balanced on baseline allowable factors. Due to the subsequent randomization step, all intervention arms share the same balance of baseline allowable factors across social groups, so that disparity is measured similarly across arms. The randomization step eliminates also systematic differences between intervention arms, such that the estimand defines meaningful intervention effects on that measure of disparity when allowability matters.

A third class of estimands are those of causal decomposition analysis,[42-44] where no intervention is made to change who belongs to which group, and the intervention of interest is to remove the disparity in a determinant of the outcome, which can be specified in many ways. The estimand of Jackson (2021)[43] is similar to that of TS+TT in that it also considers effects on disparity in a study sample where social groups are balanced on baseline allowable factors. The key distinction of TS+TT is that it does not limit itself to interventions that completely remove disparity in an outcome determinant. In fact, in our motivating example, some intervention arms were specified to only partially remove disparity in the determinant, or to increase disparity in the determinant. It is also capable of formulating quite specific and detailed intervention strategies. TS+TT is therefore a general framework that encompasses causal decomposition analysis as a special case (i.e., interventions equalize an outcome determinant). Furthermore, one can also view the population impact-based estimands as a special case (i.e., no covariates are deemed allowable).

## Practical Considerations in Implementing the TS+TT Framework

The TS+TT framework is designed to serve as a conceptual guiding tool for evaluating the effects of interventions on disparity. It addresses important limitations of existing methods, as extensively discussed in this paper, by integrating ethical and transparent disparity definition with robust causal inference for evaluating the effects of interventions on disparity. The conceptual framework itself offers methodological scaffolding that guides study design and ensures that causal estimands are theoretically identifiable.

However, as with the TT framework, implementing TS+TT using real-world data (e.g., EMR) is subject to data-related constraints that are inherent to observational studies. For example, achieving the ideal trial-like state requires data with high quality, richness, and completeness to meet foundational assumptions such as exchangeability, positivity, and consistency. These data challenges should not be seen as shortcomings of the framework itself, but rather as obstacles related to its practical application. When the core elements of TS+TT protocol (e.g., well-defined interventions, sufficient covariates) are fully met, the framework can produce valid causal estimates and effectively emulate an ideal trial design. That said, when data quality falls short—due to factors such as unmeasured confounders, sparse covariate coverage, or missingness—the results should be interpreted with due caution. These limitations are inherent to the observational nature of EMR data and are not unique to TS+TT; they reflect broader challenges in causal inference.[45,46]

To optimize the application of TS+TT in the presence of data-related constraints, researchers should adopt strategies to improve data quality and rigorously address potential biases. Combining EMR data with external sources such as registries, clinical trials, or administrative records may help fill gaps in covariate coverage and reduce measurement error. If possible, conducting sensitivity analyses to quantify the potential impact of violations in key assumptions, such as consistency and positivity, can further strengthen the reliability of causal estimates.